\documentclass[aps,prd,preprint,preprintnumbers,unsortedaddress,superscriptaddress,showpacs,nofootinbib]{revtex4-1}

\usepackage{graphicx,color}
\usepackage{relsize}
\usepackage{slashed}
\usepackage{color}
\usepackage{ulem}
\usepackage{tabu}
\usepackage{bm}
\usepackage{amsmath}
\usepackage{amssymb}
\usepackage{amsthm}
\usepackage{mathtools}
\usepackage{mathrsfs}
\usepackage{graphicx}
\usepackage{epstopdf}
\usepackage{fancyhdr}
\usepackage{array}
\usepackage[all]{xy}
\usepackage{euscript}
\usepackage{enumerate}
\usepackage{slashed}
\usepackage{hyperref}
\usepackage{subfigure} 
\usepackage{epstopdf} 
\usepackage[utf8]{inputenc}
\usepackage{cancel}

\hypersetup{colorlinks=true,linkcolor=blue,citecolor=blue,menucolor=black,urlcolor=blue,filecolor=blue}

\newcommand{\tr}{{\rm Tr}}

\newcommand{\gev}{{\rm GeV}}
\newcommand{\mev}{{\rm MeV}}
\newcommand{\cs}{{\rm Case}}

\newcommand{\erpm}[3]{{#1}^{+ #2}_{- #3}}
\newcommand{\errpm}[3]{{#1}^{#2}_{#3}}
\newcommand{\mL}{\mathcal{L}}

\newcommand{\mA}{\mathcal{A}}
\newcommand{\mC}{\mathcal{C}}

\renewcommand\sout{\bgroup \color[rgb]{1,0,0} \ULdepth=-.5ex \ULset}

\newcommand{\itp}{\affiliation{CAS Key Laboratory of Theoretical Physics, Institute of Theoretical Physics, Chinese Academy of Sciences,  Zhong Guan Cun East Street 55, Beijing 100190, China}}

\newcommand{\ucas}{\affiliation{School of Physical Sciences, University of Chinese Academy of Sciences, Beijing 100049, China}}

\begin{document}

\title{Decays of $P_c$ into $J/\psi N$ and $\eta_cN$ with heavy quark spin symmetry}

\author{Shuntaro Sakai}
\email{shsakai@itp.ac.cn}
\itp
\author{Hao-Jie Jing}
\email{jinghaojie@itp.ac.cn}
\itp\ucas
\author{Feng-Kun Guo}
\email{fkguo@itp.ac.cn}
\itp\ucas

\begin{abstract}
We investigate the consequences of heavy quark spin symmetry (HQSS) on hidden-charm pentaquark $P_c$ states.
As has been proposed before, assuming the $P_c(4440)$ and the $P_c(4457)$ as $S$-wave $\bar{D}^*\Sigma_c$ molecules, 
seven hadronic molecular states composed of $\bar{D}\Sigma_c$, $\bar{D}\Sigma_c^*$, $\bar{D}^*\Sigma_c$, and $\bar{D}^*\Sigma_c^*$ can be obtained, with the $\bar{D}\Sigma_c$ molecule corresponding to the $P_c(4312)$. 
These seven states can decay into $J/\psi N$ and $\eta_c N$, and we use HQSS to predict ratios of partial widths of the $S$-wave decays.
For the decays into $J/\psi N$, it is found that among all six $P_c$ molecules with spin $1/2$ or $3/2$, at least four states decay much more easily into the $J/\psi N$ than the $P_c(4312)$, and two of them couple dominantly to the $\bar D^*\Sigma_c^*$. While no significant peak around the $\bar{D}^*\Sigma_c^*$ threshold is found in the $J/\psi p$ distribution, these higher $P_c$ states either are produced with lower rates or some special production mechanism for the observed $P_c$ states might play an important role, such as an intricate interplay between the production of pentaquarks and triangle singularities.

\end{abstract}

\maketitle

\section{Introduction}
\label{sec:intro}
Following the first observation of two hidden-charm pentaquark candidates, the $P_c$ states named $P_c(4380)$ and $P_c(4450)$, in the $J/\psi p$ invariant mass distribution 
of the decay $\Lambda_b\rightarrow J/\psi K^-p$~\cite{Aaij:2015tga},
the LHCb Collaboration reported three narrow peaks $P_c(4312)$, $P_c(4440)$, and $P_c(4457)$ in Ref.~\cite{Aaij:2019vzc} with the full run I and run II datasets. The $P_c(4312)$ peak is new (it seems to stick out in the background in a single bin in the coarser binning in Ref.~\cite{Aaij:2015tga}), and the $P_c(4450)$ structure is split into two finer narrow peaks, $P_c(4440)$ and $P_c(4457)$.
Because these peaks are close to the $\bar{D}\Sigma_c$ and $\bar{D}^*\Sigma_c$ thresholds, the
interpretation of them as $\bar{D}^{(*)}\Sigma_c$ hadronic molecules is a natural idea.
A hadronic molecule is a bound state of two color-singlet hadrons. Analogous to the deuteron and other nuclei as proton-neutron bound states, hadronic molecules are expected to provide rich structure in the hadron spectrum (for a review of hadronic molecules, we refer to Ref.~\cite{Guo:2017jvc}).
One famous example of hadronic molecular candidates in the light baryon sector is the $\Lambda(1405)$, which is well described as an $S$-wave $\bar{K}N$ bound state~\cite{Dalitz:1967fp,Kaiser:1995eg,Oset:1997it,Oller:2000fj} 
(see Refs.~\cite{Hyodo:2011ur,Kamiya:2016jqc} for review articles). 
Hadronic molecular pentaquarks with a hidden charm were expected to exist~\cite{Wu:2010jy,Wu:2010vk,Wang:2011rga,Yang:2011wz,Yuan:2012wz,Wu:2012md,Xiao:2013yca,Uchino:2015uha,Karliner:2015ina} prior to the LHCb discovery.
The new observation of the three peaks in the $J/\psi p$ channel~\cite{Aaij:2019vzc} has been particularly encouraging in studies in this field~\cite{Xiao:2019mst,Xiao:2019aya,Liu:2019tjn,Chen:2019bip,Chen:2019asm,Fernandez-Ramirez:2019koa,Guo:2019fdo,He:2019ify,Zhu:2019iwm,Huang:2019jlf,Ali:2019npk,Shimizu:2019ptd,Guo:2019kdc,Mutuk:2019snd,Weng:2019ynv,Eides:2019tgv,Wang:2018waa,Wang:2019got,Meng:2019ilv,Cheng:2019obk,Wang:2019nwt,Wu:2019rog,Wang:2019krd,Cao:2019kst,Ali:2019lzf,Wang:2019dsi,Wu:2019adv,Holma:2019lxe,Yamaguchi:2019seo}
(see also Refs.~\cite{Chen:2016qju,Lebed:2016hpi,Esposito:2016noz,Ali:2017jda,Guo:2017jvc,Olsen:2017bmm,Karliner:2017qhf,Cerri:2018ypt,Liu:2019zoy} for reviews of the earlier literature).

For the study of hadronic systems containing heavy quarks, heavy quark spin symmetry (HQSS), 
which emerges because of the decoupling of the heavy quark spin in the limit of an infinitely large quark mass in the Lagrangian of quantum chromodynamics (QCD)~\cite{Isgur:1991wq,Wise:1992hn,Neubert:1993mb}, 
is an essential tool for making predictions. Different scenarios of the exotic hadrons are expected to lead to HQSS predictions that can be used to distinguish them~\cite{Cleven:2015era}.
Particularly in the $P_c$ mass region, there exist $\bar{D}\Sigma_c$, $\bar{D}\Sigma_c^*$, and $\bar{D}^*\Sigma_c$ thresholds. 
Since the $(\bar{D},\bar{D}^*)$ and $(\Sigma_c,\Sigma_c^*)$ pairs can be settled into HQSS doublets, respectively, it is natural to investigate the $\bar{D}^{(*)}\Sigma_c^{(*)}$ systems together by using HQSS. 
The pioneering work using HQSS to predict hidden-charm pentaquarks is Ref.~\cite{Xiao:2013yca} before the $P_c$ discovery, which was extended to the hidden-charm strange sector recently~\cite{Xiao:2019gjd}. After the $P_c(4450)$ discovery, HQSS was used in Ref.~\cite{Liu:2018zzu} to predict $\bar{D}^{(*)}\Sigma_c^{(*)}$ molecules, and the results were updated after the new LHCb observation in Ref.~\cite{Liu:2019tjn}.
In Ref.~\cite{Liu:2018zzu}, the nonrelativistic contact term Lagrangian for the $S$-wave $\bar{D}^{(*)}\Sigma_c^{(*)}$ interaction respecting HQSS is constructed, 
which was used in Ref.~\cite{Liu:2019tjn} to predict a whole set of seven states related to each another via HQSS. The states are generated in an $S$-wave by the following channels: $\bar D\Sigma_c (1/2^-)$, $\bar D\Sigma_c^* (3/2^-)$, $\bar D^*\Sigma_c (1/2^-,3/2^-)$, and $\bar D^*\Sigma_c^* (1/2^-, 3/2^-,5/2^-)$. The predictions were made by fixing the only two parameters to reproduce the masses of the $P_c(4440)$ and $P_c(4457)$ as $J^P=1/2^-$ and $3/2^-$ $\bar{D}^*\Sigma_c$ molecular states. The results  obtained in Ref.~\cite{Xiao:2019aya} using a  different formalism are similar, and the reference also finds good agreement between their results and measured values for the widths of observed $P_c$ peaks.

On the one hand, in the updated LHCb measurements~\cite{Aaij:2019vzc}, the $P_c(4312)$ is discovered with a significance of $7.3\sigma$, and the most visible structure, at around 4.45~GeV, is resolved into two narrow peaks, $P_c(4440)$ and $P_c(4457)$, with a significance of $5.4\sigma$, while there are no other peaking structures that can be unambiguously distinguished from statistical fluctuations. On the other hand, in the hadronic molecular picture, seven states are expected to exist with six of them being able to decay into the $J/\psi N$ in an $S$-wave. Therefore, one important question to be answered in the hadronic molecular model is why only three $P_c$ states were observed. To answer this question, the decays of the $P_c$ states into the $J/\psi p$ are an essential ingredient.

In Ref.~\cite{Xiao:2019mst}, the decays of the observed three $P_c$ states into $J/\psi p$ through the $\bar{D}^{(*)}\Sigma_c^{(*)}D^{(*)}$ triangle loops are considered, and the obtained partial widths are of the order of a few to 10~MeV. Since the total widths of the  $P_c(4312)$, $P_c(4440)$ and $P_c(4457)$ are $9.8_{-5.2}^{+4.6}$~MeV, $20.6_{-11.2}^{+10.0}$~MeV, and $6.4_{-2.8}^{+6.0}$~MeV, respectively,\footnote{The statistical and systematic uncertainties in Ref.~\cite{Aaij:2019vzc} are added in quadrature here.} the results in Ref.~\cite{Xiao:2019mst} would mean that the branching fractions of the $J/\psi N$ mode are much larger than the model-dependent upper limit set by the GlueX experiment~\cite{Ali:2019lzf}.\footnote{The results in Refs.~\cite{Shen:2016tzq,Lin:2017mtz,Lin:2019qiv} indicate that the dominant decay modes of the $P_c$ states should be $\bar D^{(*)}\Lambda_c$ instead of $J/\psi N$.}  One notices, however, that the partial widths obtained in Ref.~\cite{Xiao:2019mst} depend on unknown couplings (for the $J/\psi D\bar D$, the $\Sigma_c D N$, and their HQSS related vertices) and are sensitive to the cutoff value introduced to regularize the ultraviolet divergent triangle loop integrals. The decays of the $1/2^-$ $P_c$ states as $D^{(*)}\Sigma_c$ hadronic molecules into $J/\psi N,\eta_c N$ and $\bar D^{(*)}\Lambda_c$ were very recently discussed in Ref.~\cite{Voloshin:2019aut} by considering HQSS.

In this paper, we investigate the decays of all six $P_c$ hadronic molecules with $J^P=1/2^-$ or $3/2^-$ into $J/\psi N$ and $\eta_cN$ with a formulation respecting HQSS, and we predict ratios of the partial widths, which are free of unknown coupling constants. The paper is organized as follows. The amplitudes are worked out in Section~\ref{sec:form} with details given in Appendix~\ref{app:interaction}. Numerical results and related discussions are presented in Section~\ref{sec:res}. Section~\ref{sec:sum} is a brief summary.

\section{Formalism}
\label{sec:form}

In this section, we describe the $P_c$ states as hadronic molecules  which are dynamically generated 
from the $\bar{D}^{(*)}\Sigma_c^{(*)}$ $S$-wave short-range interactions respecting HQSS.
The transition amplitudes for $\bar{D}^{(*)}\Sigma_c^{(*)}$ into $J/\psi N$ and $\eta_c N$ will also be constructed.

\subsection{Short-range \texorpdfstring{$\bm{\bar{D}^{(*)}\Sigma_c^{(*)}}$}{Dbar Sigmac} interactions}

To describe the $\bar{D}^{(*)}\Sigma_c^{(*)}$ molecular $P_c$ states, we start with the interaction respecting HQSS. Here, following Refs.~\cite{Liu:2018zzu,Liu:2019tjn}, we consider the short-range coupled-channel interactions\footnote{Channel couplings are not considered in Refs.~\cite{Liu:2018zzu,Liu:2019tjn}.} which can be parametrized in terms of contact terms. As a consequence of HQSS, for each total isospin (here $I=1/2$), all possible $S$-wave short-range $\bar{D}^{(*)}\Sigma_c^{(*)}$ interactions at leading order (LO) of the nonrelativistic expansion depend on only two parameters. 
The LO potentials for the system with total spin $J$ can be easily worked out by using either the $9j$ symbol as in Refs.~\cite{Xiao:2013yca,Xiao:2019aya} or by constructing the LO effective Lagrangian as in Refs.~\cite{Liu:2018zzu,Liu:2019tjn}, and the details can be found in Appendix~\ref{app:interaction}. 
They are given by
\begin{align}
 v_{X,X'(J)}=C_a\,c_{X,X'(J)}+C_b\,c'_{X,X'(J)}~~~(X^{(\prime)}=\bar{D}\Sigma_c,\bar{D}\Sigma_c^*,\bar{D}^*\Sigma_c,\bar{D}^*\Sigma_c^*;J=\frac12,\frac32,\frac52),
 \label{eq:v}
\end{align}
where $C_a$ and $C_b$ are energy-independent constants, and $c_{X,X'(J)}$ and $c'_{X,X'(J)}$ are coefficients which depend on the channels of the initial and final states as tabulated in Table~\ref{table:CG}.
\begin{table}[t]
 \centering
 \caption{The contact terms for the coupled-channel $\bar D^{(*)}\Sigma_c^{(*)}$ interactions for $J=1/2$ (left panel), $J=3/2$ (middle panel), and $J=5/2$ (right panel).}
 \label{table:CG}
\begin{minipage}[t]{0.35\hsize}
  \begin{tabular}[t]{c|ccc}
  \hline\hline
 $J=\frac12$ &$\bar{D}\Sigma_c$ &$\bar{D}^*\Sigma_c$ &$\bar{D}^*\Sigma_c^*$ \\\hline
  $\bar{D}\Sigma_c$&$C_a$ &$\frac{2}{\sqrt{3}}C_b$ &$-\sqrt{\frac{2}{3}}C_b$ \\
  $\bar{D}^*\Sigma_c$& $\frac{2}{\sqrt{3}}C_b$ &$C_a-\frac{4}{3}C_b$ &$-\frac{\sqrt{2}}{3}C_b$ \\
  $\bar{D}^*\Sigma_c^*$&$-\sqrt{\frac{2}{3}}C_b$  & $-\frac{\sqrt{2}}{3}C_b$ & $C_a-\frac{5}{3}C_b$\\
  \hline\hline
 \end{tabular}
\end{minipage} \hfill
\begin{minipage}[t]{0.35\hsize}
  \begin{tabular}[t]{c|ccc}
  \hline\hline
 $J=\frac32$ &$\bar{D}\Sigma_c^*$ &$\bar{D}^*\Sigma_c$ &$\bar{D}^*\Sigma_c^*$ \\\hline
  $\bar{D}\Sigma_c^*$&$C_a$ &$\frac{1}{\sqrt{3}}C_b$ &$\sqrt{\frac{5}{3}}C_b$ \\
  $\bar{D}^*\Sigma_c$& $\frac{1}{\sqrt{3}}C_b$ &$C_a+\frac{2}{3}C_b$ &$-\frac{\sqrt{5}}{3}C_b$ \\
  $\bar{D}^*\Sigma_c^*$& $\sqrt{\frac{5}{3}}C_b$ & $-\frac{\sqrt{5}}{3}C_b$ & $C_a-\frac{2}{3}C_b$\\
  \hline\hline
 \end{tabular}
\end{minipage} \hfill
\begin{minipage}[t]{0.15\hsize}
  \begin{tabular}[t]{c|c}
  \hline\hline
  $J=\frac52$ &$\bar{D}^*\Sigma_c^*$ \\\hline
  $\bar{D}^*\Sigma_c^*$& $C_a+C_b$\\
  \hline\hline
 \end{tabular}
\end{minipage}
\end{table}
As one can see, the diagonal potentials depend on both $C_a$ and $C_b$, while the channel coupling is controlled by the parameter $C_b$.

The $T$-matrix of the $\bar{D}^{(*)}\Sigma_c^{(*)}$ scattering, $t$, is obtained by resumming the $s$-channel bubbles with the coupled-channel Lippmann-Schwinger equation, which satisfies unitarity,
\begin{align}
 t=[1-v\, G]^{-1}v,\label{eq:SLeq}
\end{align}
where $v$ in Eq.~\eqref{eq:v} is used
as the interaction kernel,\footnote{There is a factor from the nonrelativistic normalization of the heavy meson fields; see Appendix~\ref{app:interaction}.} and $G$ is a diagonal matrix  given by the nonrelativistic meson-baryon loop functions.
Using a Gaussian form factor $f(q/\Lambda)=e^{-q^2/\Lambda^2}$ to regularize the ultraviolet divergence as in Refs.~\cite{Liu:2018zzu,Liu:2019tjn}, 
the loop function $G_X(W)$ in channel $X$ as a function of the total energy, $W$, in the meson-baryon center-of-mass (CM) frame is given by
\begin{align}
 G_X(W)=\frac{2M_X}{4m_XM_X}\int\frac{d^3q}{(2\pi)^3}\frac{e^{-2q^2/\Lambda^2}}{W-m_X-M_X-q^2/(2\mu_X)+i\epsilon}, \label{eq:G}
\end{align}
where $m_X$ and $M_X$ denote the meson and baryon masses in that channel, respectively, and $\mu_X=m_XM_X/(m_X+M_X)$ is the meson-baryon reduced mass.
In this work, 
we take isospin averaged hadron masses,
and the $\bar{D}^{(*)}$ and $\Sigma_c^{(*)}$ widths are ignored.\footnote{The widths of the $\Sigma_c^*$ states are around 15~MeV~\cite{Tanabashi:2018oca}, similar to the measured widths of the $P_c$ states. The decays of the $\Sigma_c^*\bar D^{(*)}$ molecules through the decays of the $\Sigma_c^*$ into $\Lambda_c\pi$ might contribute an important portion of the total widths of these states.}
Two values of the Gaussian cutoff $\Lambda$,  $0.7~\gev$ and $1~\gev$, will be taken in order to check the uncertainty of the results. The values are chosen such that they are larger than the binding momenta in all of the involved channels (much larger than that in the dominant one) and still much smaller than the charmed hadron masses so that no significant HQSS breaking will be introduced by $\Lambda$. 
The $T$-matrix in Eq.~\eqref{eq:SLeq} has poles, and the real parts correspond to the masses of the hadronic molecules generated from the interactions.

In Eq.~\eqref{eq:v}, there are two constants, $C_a$ and $C_b$, that should be determined.
We fix these two parameters so as to reproduce the observed peak positions of the $P_c(4440)$ and the $P_c(4457)$.
Following Ref.~\cite{Liu:2019tjn}, we consider two cases for the spin assignment of $P_c(4440)$ and $P_c(4457)$ as $\bar{D}^*\Sigma_c$ molecular states:
\begin{itemize}
 \item[]Case~1: $P_c(4440)$ and $P_c(4457)$ have $J=1/2$ and $3/2$, respectively;
 \item[]Case~2: $P_c(4440)$ and $P_c(4457)$ have $J=3/2$ and $1/2$, respectively.
\end{itemize}
The parameters $C_a$ and $C_b$ in these two cases are given in the left and right panels of Table~\ref{tab_prmtcab}, respectively.
\begin{table}[t]
 \centering
   \caption{Parameters $C_a$ and $C_b$ fixed in Case~1 (left panel) and Case~2 (right panel) with $\Lambda=0.7~\gev$ and $1~\gev$.}
   \label{tab_prmtcab}
 \begin{minipage}[c]{0.47\hsize}
  \begin{tabular}[t]{c|c|c}
  \hline\hline
   $\Lambda~[\gev]$&$C_a~[\gev^{-2}]$ &$C_b~[\gev^{-2}]$ \\\hline
   $0.7$&$-33.0$ &$6.6$ \\
   $1$&$-20.0$ &$2.9$\\
   \hline\hline
  \end{tabular}
 \end{minipage}
 \hfill
 \begin{minipage}[c]{0.47\hsize}
    \begin{tabular}[t]{c|c|c}
  \hline\hline
   $\Lambda~[\gev]$&$C_a~[\gev^{-2}]$ &$C_b~[\gev^{-2}]$ \\\hline
   $0.7$&$-37.2$ &$-5.8$ \\
   $1$&$-21.9$ &$-2.6$\\
  \hline\hline
  \end{tabular}
 \end{minipage}
\end{table}
In both two cases, the magnitude of $C_b$ is much smaller than that of $C_a$ in order to produce poles at $4440~\mev$ and $4457~\mev$ in the $\bar{D}^*\Sigma_c$ channel. From Table~\ref{table:CG}, this means that the channel coupling is rather weak, and all diagonal interactions have similar strengths so that one expects to have seven  $\bar{D}^{(*)}\Sigma_c^{(*)}$ hadronic molecules.

By choosing appropriate Riemann sheets we find resonance and bound-state poles.
As in Refs.~\cite{Liu:2019tjn,Xiao:2019aya}, seven states of $\bar{D}\Sigma_c(J=1/2)$, $\bar{D}\Sigma_c^*(J=3/2)$, $\bar{D}^*\Sigma_c(J=1/2,3/2)$, and $\bar{D}^*\Sigma_c^*(J=1/2,3/2,5/2)$ are obtained as a consequence of HQSS.
These seven states are denoted by $P_{ci}~(i=1\sim 7)$, and their pole positions for Case 1 and Case 2 are listed in Tables~\ref{tab:pole1} and \ref{tab:pole2}, respectively. 
For each of these states, its effective coupling constants to the meson-baryon channels can be obtained from the residues of the corresponding pole of the $T$-matrix elements, namely 
\begin{equation}
    g_\text{eff}^i\, g_\text{eff}^j = \lim _{W\to W_\text{pole}} (W-W_\text{pole})\, t^{ij}(W)\,.
    \label{eq:geff}
\end{equation}
The so-obtained effective coupling constants are given in Tables~\ref{tab:coupling1} and \ref{tab:coupling2} for Case 1, and in Tables~\ref{tab:coupling3} and \ref{tab:coupling4} for Case 2.
As expected from $|C_a|\gg |C_b|$, each pole couples dominantly to a single channel. 
The binding energies defined as the difference between the threshold of the dominant channel and the real part of the pole are also listed in Tables~\ref{tab:pole1} and \ref{tab:pole2}.
For each spin $J$, the lowest state is a bound-state pole, while the higher ones are resonance poles ($P_{c3,c4,c5,c6}$) with a small imaginary part, which is again due to the smallness of $C_b$ which appears in the off-diagonal part of the interaction kernel. The absolute value of the imaginary part can be identified as half of the partial width of the decays of that state into the channels with lower thresholds.

\begin{table}[tp]
 \caption{Poles in units of $\mev$ in Case~1.
 The channel which has the largest coupling is given in the second column.
 The binding energies with respect to that channel are shown in parentheses in the last two columns.
 {The uncertainties of the $P_{c1,2,5,6,7}$ poles are evaluated by changing the parameters $C_a$ and $C_b$ by $25\%$.
 The poles marked with ``(V)'' move into a wrong Riemann sheet that is not directly connected to the physical region by crossing the cut at the energy of the real part.
 The poles $P_{c2}$ and $P_{c3}$ are used as input to fix the parameters and thus do not have such an uncertainty.}}
 \label{tab:pole1}
 \vspace{-3mm}
 \begin{ruledtabular}
 \begin{tabular}[t]{c|ccc}
  &Dominant channel &$\Lambda=0.7~\gev$ &$\Lambda=1~\gev$ \\\hline
  $P_{c1}$&$\bar{D}\Sigma_c~(J=1/2)$&$\erpm{4311.1}{8.5}{13.5}$ $(\erpm{9.7}{13.5}{8.5})$ &$\erpm{4311.8}{8.9}{19.2}$ $(\erpm{8.9}{19.2}{8.9})$ \\
  $P_{c2}$&$\bar{D}\Sigma_c^*~(J=3/2)$&$\erpm{4374.9}{9.0}{14.2}$ $(\erpm{10.5}{14.2}{9.0})$ &$\erpm{4375.5}{9.7}{20.3}$ $(\erpm{9.9}{20.3}{9.7})$ \\
  $P_{c3}$&$\bar{D}^*\Sigma_c~(J=1/2)$&$4440.3-i1.0$ $(21.8)$ &$4440.3-i1.2$ $(21.8)$\\
  $P_{c4}$&$\bar{D}^*\Sigma_c~(J=3/2)$&$4457.3-i0.5$ $(4.8)$ &$4457.3-i0.5$ $(4.8)$\\
  $P_{c5}$&$\bar{D}^*\Sigma_c^*~(J=1/2)$&$\erpm{4501.0}{16.4}{19.8}+i\erpm{(-0.9)}{0.5}{0.7}$ $(\erpm{25.7}{19.8}{16.4})$ &$\erpm{4500.6}{20.3}{27.9}+i\erpm{(-1.1)}{0.7}{0.8}$ $
(\erpm{26.0}{27.9}{20.3})$\\
  $P_{c6}$&$\bar{D}^*\Sigma_c^*~(J=3/2)$&$\erpm{4513.1}{10.2}{13.6}+i\erpm{(-1.9)}{1.1}{1.1}$ $(\erpm{13.6}{13.6}{10.2})$ &$\erpm{4512.9}{12.4}{19.9}+i\erpm{(-2.2)}{1.6}{1.4}$ $
(\erpm{13.8}{19.9}{12.4})$\\
  $P_{c7}$&$\bar{D}^*\Sigma_c^*~(J=5/2)$&$\errpm{4523.9}{\text{(V)}}{-11.6}$ $(\errpm{2.8}{+11.6}{\text{(V)}})$ &$\errpm{4523.8}{\text{(V)}}{-16.3}$ $(\errpm{2.9}{+16.3}{\text{(V)}})$
 \end{tabular}
 \end{ruledtabular}
\end{table}

\begin{table}[tbp]
 \caption{Poles in units of $\mev$ in Case~2.
 The channel which has the largest coupling is given in the second column.
 The binding energies with respect to that channel are shown in parentheses in the last two columns.
 {The errors are the same as those in Table~\ref{tab:pole1}.}}
 \label{tab:pole2}
 \vspace{-3mm}
  \begin{ruledtabular}
 \begin{tabular}[t]{c|ccc}
  &Dominant channel &$\Lambda=0.7~\gev$ &$\Lambda=1~\gev$ \\\hline
  $P_{c1}$&$\bar{D}\Sigma_c~(J=1/2)$&$\erpm{4305.9}{11.5}{16.1}$ $(\erpm{14.8}{16.1}{11.5})$ &$\erpm{4306.8}{12.9}{22.5}$ $(\erpm{13.9}{22.5}{12.9})$\\
  $P_{c2}$&$\bar{D}\Sigma_c^*~(J=3/2)$&$\erpm{4369.8}{11.9}{16.5}$ $(\erpm{15.5}{16.5}{11.9})$ &$\erpm{4370.5}{13.7}{23.3}$ $(\erpm{14.9}{23.3}{13.7})$\\
  $P_{c3}$&$\bar{D}^*\Sigma_c~(J=1/2)$&$4457.3-i0.5$ $(4.8)$ &$4457.3-i0.6$ $(4.8)$\\
  $P_{c4}$&$\bar{D}^*\Sigma_c~(J=3/2)$&$4440.3-i0.2$ $(21.8)$ &$4440.3-i0.2$ $(21.8)$\\
  $P_{c5}$&$\bar{D}^*\Sigma_c^*~(J=1/2)$&$\errpm{4523.3}{\text{(V)}}{-15.1}+i\errpm{(-0.2)}{+0.2}{-0.2}$ $(\errpm{3.3}{+15.1}{\text{(V)}})$ &$\errpm{4523.2}{\text{(V)}}{-20.3}+i\errpm{(-0.3)}{+0.3}{-0.3}$ $(\errpm{\
3.5}{+20.3}{\text{(V)}})$\\
  $P_{c6}$&$\bar{D}^*\Sigma_c^*~(J=3/2)$&$\errpm{4518.1}{\text{(V)}}{-16.3}+i\errpm{(-1.2)}{+0.8}{-0.7}$ $(\errpm{8.6}{+16.3}{\text{(V)}})$ &$\errpm{4517.9}{\text{(V)}}{-22.2}+i\errpm{(-1.4)}{+1.2}{-1.0}$ $(\errpm{8.8}{+22.2}{\text{(V)}})$\\
  $P_{c7}$&$\bar{D}^*\Sigma_c^*~(J=5/2)$&$\erpm{4501.6}{16.2}{20.1}$ $(\erpm{25.0}{20.1}{16.2})$ &$\erpm{4501.3}{20.0}{28.3}$ $(\erpm{25.3}{28.3}{20.0})$
 \end{tabular}
  \end{ruledtabular}
\end{table}

{It is important to understand the robustness of the predictions against the breaking of HQSS. Possible uncertainties of the predicted pole positions from the higher-order correction of the $1/m_Q$ expansion, where $m_Q$ denotes the heavy quark mass, are conservatively estimated by changing the low-energy constants, $C_a$ and $C_b$, by an amount of $\Lambda_{\rm QCD}/m_c\simeq25\%$.
It is noticeable that $P_{c1}$, which is associated with $P_c(4312)$, and $P_{c2}$, a $\bar{D}\Sigma_c^*$ molecule,  are stable in this range of uncertainty.
Furthermore, at least one or two of the three poles around the $\bar{D}^*\Sigma_c^*$ threshold, $P_{c5,6,7}$, remain even if one changes the parameters $C_a$ and $C_b$ by $25\%\
$. Within the uncertainties, the other poles may move into a ``wrong'' Riemann sheet that is not directly connected to the physical region by crossing the cut at the energy of the real part (the physical region can be reached by bypassing the threshold branching point; see, e.g., Ref.~\cite{Guo:2006fu}, and see also a recent analysis of the $P_c(4312)$ in Ref.~\cite{Fernandez-Ramirez:2019koa}). Such situations are marked ``(V)'', meaning virtual state, in the tables. In that case, the poles are still close to the threshold,
and they can still show up in invariant mass distributions as a peak with a pronounced cusp structure at the $\bar{D}^*\Sigma_c^*$ threshold. For simplicity, in the following discussions of decays, we will neglect the uncertainties and keep in mind that the results are obtained assuming exact HQSS for the interaction vertices.}

\begin{table}[t]
 \caption{Coupling constants in Case~1 with $\Lambda=0.7~\gev$. The coupling constants are dimensionless}
 \label{tab:coupling1}
 \vspace{-3mm}
 \begin{ruledtabular}
 \begin{tabular}[t]{c|cccc}
  &$\bar{D}\Sigma_c$ &$\bar{D}\Sigma_c^*$ &$\bar{D}^*\Sigma_c$ &$\bar{D}^*\Sigma_c^*$ \\\hline
  $P_{c1}~(J=1/2)$&$2.34$ &$-$ &$-0.90$ &$0.54$ \\ 
  $P_{c2}~(J=3/2)$&$-$ &$2.39$ &$-0.53$ &$-1.03$ \\ 
  $P_{c3}~(J=1/2)$&$-0.26-i0.24$ &$-$ &$3.34+i0.09$ &$0.75-i0.10$ \\ 
  $P_{c4}~(J=3/2)$&$-$ &$-0.11-i0.19$ &$1.82+i0.06$ &$0.60+i0.12$ \\ 
  $P_{c5}~(J=1/2)$&$0.28+i0.12$ &$-$ &$0.03+i0.20$ &$3.61+i0.07$ \\ 
  $P_{c6}~(J=3/2)$&$-$ &$-0.28-i0.26$ &$0.10+i0.26$ &$2.71+i0.18$ \\ 
  $P_{c7}~(J=5/2)$&$-$ &$-$ &$-$ &$1.50$ 
 \end{tabular}
 \end{ruledtabular}
\end{table}

\begin{table}[t]
 \caption{Coupling constants in Case~1 with $\Lambda=1~\gev$.
 The coupling constants are dimensionless.}
 \label{tab:coupling2}
 \vspace{-3mm}
 \begin{ruledtabular}
 \begin{tabular}[t]{c|cccc}
  &$\bar{D}\Sigma_c$ &$\bar{D}\Sigma_c^*$ &$\bar{D}^*\Sigma_c$ &$\bar{D}^*\Sigma_c^*$ \\\hline
  $P_{c1}~(J=1/2)$&$2.08$ &$-$ &$-0.69$ &$0.40$ \\ 
  $P_{c2}~(J=3/2)$&$-$ &$2.14$ &$-0.42$ &$-0.79$ \\ 
  $P_{c3}~(J=1/2)$&$-0.15-i0.26$ &$-$ &$2.94+i0.08$ &$0.61-i0.09$ \\ 
  $P_{c4}~(J=3/2)$&$-$ &$-0.05-i0.19$ &$1.70+i0.06$ &$0.49+i0.11$ \\ 
  $P_{c5}~(J=1/2)$&$0.18+i0.16$ &$-$ &$-0.003+i0.19$ &$3.16+i0.07$ \\ 
  $P_{c6}~(J=3/2)$&$-$ &$-0.15-i0.28$ &$0.03+i0.25$ &$2.45+i0.16$ \\ 
  $P_{c7}~(J=5/2)$&$-$ &$-$ &$-$ &$1.44$ 
 \end{tabular}
 \end{ruledtabular}
\end{table}

\begin{table}[t]
 \caption{Coupling constants in Case~2 with $\Lambda=0.7~\gev$.
 The coupling constants are dimensionless.}
 \label{tab:coupling3}
 \vspace{-3mm}
 \begin{ruledtabular}
 \begin{tabular}[t]{c|cccc}
  &$\bar{D}\Sigma_c$ &$\bar{D}\Sigma_c^*$ &$\bar{D}^*\Sigma_c$ &$\bar{D}^*\Sigma_c^*$ \\\hline
  $P_{c1}~(J=1/2)$& $2.79$ &$-$ &$0.73$ &$-0.48$ \\ 
  $P_{c2}~(J=3/2)$& $-$ &$2.83$ &$0.48$ &$0.81$ \\
  $P_{c3}~(J=1/2)$& $0.20+i0.17$ &$-$ &$1.83+i0.07$ &$-0.25-i0.04$ \\ 
  $P_{c4}~(J=3/2)$& $-$ &$0.04+i0.13$ &$3.35+i0.02$ &$-0.70+i0.05$ \\ 
  $P_{c5}~(J=1/2)$& $-0.13-i0.08$ &$-$ &$-0.05-i0.10$ &$1.60+i0.04$ \\ 
  $P_{c6}~(J=3/2)$& $-$ &$0.25+i0.21$ &$-0.06-i0.19$ &$2.26+i0.13$ \\ 
  $P_{c7}~(J=5/2)$& $-$ &$-$ &$-$ &$3.57$ 
 \end{tabular}
 \end{ruledtabular}
\end{table}

\begin{table}[t]
 \caption{Coupling constants in Case~2 with $\Lambda=1~\gev$.
 The coupling constants are dimensionless.}
 \label{tab:coupling4}
 \vspace{-3mm}
 \begin{ruledtabular}
 \begin{tabular}[t]{c|cccc}
  &$\bar{D}\Sigma_c$ &$\bar{D}\Sigma_c^*$ &$\bar{D}^*\Sigma_c$ &$\bar{D}^*\Sigma_c^*$ \\\hline
  $P_{c1}~(J=1/2)$&$2.45$ &$-$ &$0.56$ &$-0.36$ \\ 
  $P_{c2}~(J=3/2)$&$-$ &$2.49$ &$0.39$ &$0.62$ \\ 
  $P_{c3}~(J=1/2)$&$0.11+i0.18$ &$-$ &$1.71+i0.07$ &$-0.20-i0.04$ \\ 
  $P_{c4}~(J=3/2)$&$-$ &$0.01+i0.12$ &$2.94+i0.01$ &$-0.57+i0.04$ \\ 
  $P_{c5}~(J=1/2)$&$-0.07-i0.09$ &$-$ &$-0.02-i0.10$ &$1.53+i0.04$ \\ 
  $P_{c6}~(J=3/2)$&$-$ &$0.14+i0.23$ &$-0.01-i0.19$ &$2.09+i0.12$ \\ 
  $P_{c7}~(J=5/2)$&$-$ &$-$ &$-$ &$3.13$ 
 \end{tabular}
 \end{ruledtabular}
\end{table}

\subsection{Transition amplitudes of \texorpdfstring{$\bm{\bar{D}^{(*)}\Sigma_c^{(*)}}$}{Dbar Sigmac} into \texorpdfstring{$\bm{J/\psi N}$}{Jpsi N} and \texorpdfstring{$\bm{\eta_cN}$}{etac N}}

Next let us consider the transition amplitudes of the $\bar{D}^{(*)}\Sigma_c^{(*)}$ into the $J/\psi N$.
Using the $9j$ symbol to recombine the angular momenta (see Appendix~\ref{app:interaction}), 
we write the $S$-wave $\bar{D}^{(*)}\Sigma_c^{(*)}\rightarrow J/\psi N$ amplitude with spin $J$ as follows,
\begin{align}
 t_{X,J/\psi N(J)}=
 g_1h_{X(J)},\label{eq:tXJN}
\end{align}
where $g_1$ is a coupling constant, and
\begin{align}
 \begin{split}
  &h_{\bar{D}\Sigma_c(1/2)}=-\frac{1}{2\sqrt{3}},\quad h_{\bar{D}\Sigma_c^*(3/2)}=-\frac{1}{\sqrt{3}},\quad h_{\bar{D}^*\Sigma_c(1/2)}=\frac{5}{6},\\
  &h_{\bar{D}^*\Sigma_c(3/2)}=\frac{1}{3}, \quad
  h_{\bar{D}^*\Sigma_c^*(1/2)}=\frac{\sqrt{2}}{3},\quad h_{\bar{D}^*\Sigma_c^*(3/2)}=\frac{\sqrt{5}}{3}.
 \end{split}\label{eq:hX}
\end{align}
The $\bar D^*\Sigma_c^*$ with $J=5/2$ does not couple to the $S$-wave $J/\psi N$. One notices that all of the $S$-wave transition amplitudes depend on the same parameter $g_1$ due to HQSS. 
As a result, one can make parameter-free predictions for the ratios of partial widths.

Because the $\eta_c$ and the $J/\psi$ form a doublet of HQSS (see, e.g., Ref.~\cite{Casalbuoni:1996pg} and the references therein),
we can also relate the partial decay widths of the $P_c$ into $J/\psi N$ and $\eta_cN$.
In the same manner as the $\bar{D}^{(*)}\Sigma_c^{(*)}\rightarrow J/\psi N$ amplitude, for the $\bar{D}^{(*)}\Sigma_c^{(*)}\rightarrow\eta_cN$ one has
\begin{align}
 t_{X,\eta_c N(J=1/2)}=&\;
 g_1\tilde{h}_{X(1/2)},\label{eq:tXecN}\\
 \tilde{h}_{\bar{D}\Sigma_c(1/2)}=&\;\frac{1}{2},\quad \tilde{h}_{\bar{D}^*\Sigma_c(1/2)}=-\frac{1}{2\sqrt{3}},\quad \tilde{h}_{\bar{D}^*\Sigma_c^*(1/2)}=\sqrt{\frac{2}{3}},\label{eq:htilX}
\end{align}
where the $\eta_cN$ couples only to the states with $J=1/2$ in an $S$-wave. The ratios $\tilde{h}_{\bar{D}\Sigma_c(1/2)}/{h}_{\bar{D}\Sigma_c(1/2)}=-\sqrt{3}$ and $\tilde{h}_{\bar{D}^*\Sigma_c(1/2)}/{h}_{\bar{D}^*\Sigma_c(1/2)}=-\sqrt{3}/5$ agree with those derived in Ref.~\cite{Voloshin:2019aut}.

\subsection{\texorpdfstring{$\bm{P_c\rightarrow J/\psi N}$}{Pc to Jpsi N} and \texorpdfstring{$\bm{P_c\rightarrow\eta_cN}$}{Pc to etac N} decay amplitudes}

\begin{figure}[t]
 \centering
 \includegraphics[width=6cm]{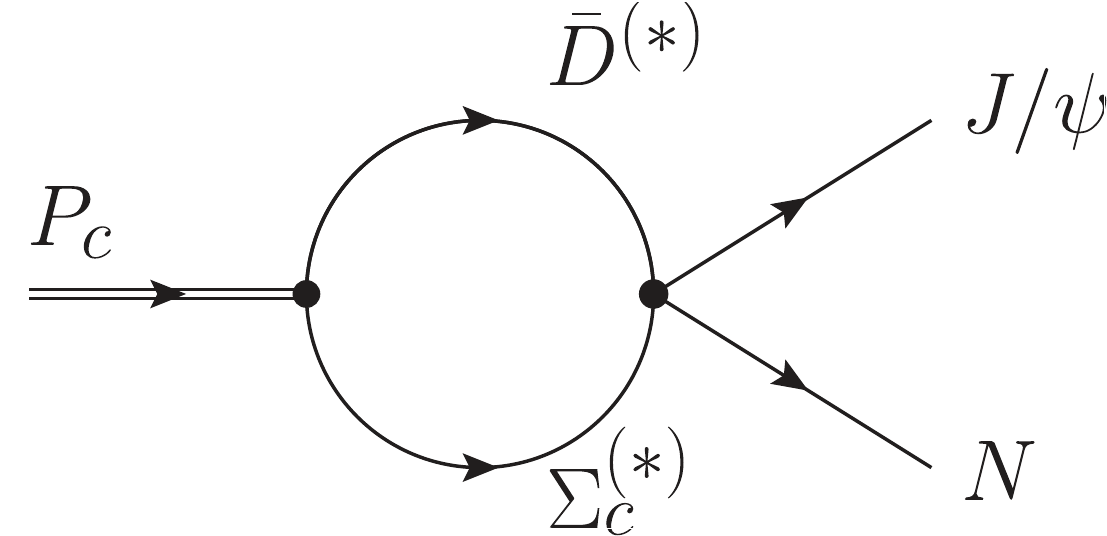}
 \caption{Diagram of the $P_{c}$ decay into $J/\psi N$ with intermediate $\bar{D}^{(*)}\Sigma_c^{(*)}$.}
 \label{fig1}
\end{figure}

The mechanism for the decay $P_c\rightarrow J/\psi N$ for the $P_c$ as $\bar D^{(*)}\Sigma_c^{(*)}$ hadronic molecules is shown in Fig.~\ref{fig1}.
The $P_{c}$ resonance first couples to $\bar{D}^{(*)}\Sigma_c^{(*)}$, and the $\bar{D}^{(*)}\Sigma_c^{(*)}$ pair turns into the $J/\psi N$ via rescattering. 
The momentum exchange for the rescattering is much larger than the binding momentum, and thus the rescattering is of short range and can be parametrized using the amplitude in Eqs.~\eqref{eq:tXJN}.\footnote{The rescattering is modeled by charmed-meson exchanges in Ref.~\cite{Xiao:2019mst}.}
The decays into $\eta_cN$ are similar.
The decay amplitudes of $P_{ci}$ with spin $J$ into the $J/\psi N$ and the $\eta_cN$, $\mA_{i(J)}$ and $\tilde{\mA}_{i(J)}$, respectively, are written as 
\begin{align}
 \mA_{i(J)}=\sum_Xg_{P_{ci},X} \tilde{G}_X g_1h_{X(J)},\label{eq:AiJ}\\
 \tilde{\mA}_{i(J)}=\sum_Xg_{P_{ci},X}\tilde{G}_X g_1\tilde{h}_{X(J)},\label{eq:AtiliJ}
\end{align}
with $X=\bar{D}\Sigma_c,\bar{D}^*\Sigma_c,\bar{D}^*\Sigma_c^*~(J=1/2)$, $X=\bar{D}\Sigma_c^*,\bar{D}^*\Sigma_c,\bar{D}^*\Sigma_c^*~(J=3/2)$, and $X=\bar{D}^*\Sigma_c^*~(J=5/2)$. Here, the coupling constants $g_{P_{ci},X}$ are those defined in Eq.~\eqref{eq:geff}.
The meson-baryon loop function in channel $X$, $\tilde{G}_X$, is given by 
\begin{align}
 \tilde{G}_X(W)=\frac{2M_X}{4m_XM_X}\int\frac{d^3q}{(2\pi)^3}\frac{e^{-q^2/\Lambda^2}}{W-m_X-M_X-q^2/(2\mu_X)+i\epsilon}, 
 \label{eq:Gtil}
\end{align}
where the Gaussian form factor $e^{-q^2/\Lambda^2}$ is introduced only for the $P_c\to \bar D^{(*)}\Sigma_c^{(*)}$ vertex, and the cutoff $\Lambda$ is chosen to be the same as that in the $\bar{D}^{(*)}\Sigma_c^{(*)}$ scattering $T$-matrix.

With these amplitudes and taking into account the nonrelativistic normalization factors, the partial decay widths are given by
\begin{align}
 \Gamma_i\equiv&\;\Gamma_{P_{ci},J/\psi N}=\frac{m_N}{2\pi m_{P_{ci}}}p_{J/\psi}|\mA_{i(J)}|^2,& p_{J/\psi}=&\;\frac{1}{2m_{P_{ci}}}\lambda^{1/2}(m_{P_{ci}}^2,m_N^2,m_{J/\psi}^2),\label{eq:Gami}\\
 \tilde{\Gamma}_i\equiv&\;\Gamma_{P_{ci},\eta_c N}=\frac{m_N}{2\pi m_{P_{ci}}}p_{\eta_c}|\tilde{\mA}_{i(J)}|^2,&p_{\eta_c}=&\;\frac{1}{2m_{P_{ci}}}\lambda^{1/2}(m_{P_{ci}}^2,m_N^2,m_{\eta_c}^2),\label{eq:Gamtili}
\end{align}
with the K\"all\'en function $\lambda(x,y,z)=x^2+y^2+z^2-2xy-2yz-2zx$. Note that the spin averaging has been taken into account in the amplitudes given by Eqs.~\eqref{eq:AiJ} and \eqref{eq:AtiliJ} (see Appendix~\ref{app:interaction}), derived using the $9j$ symbol technique, and there is no need to introduce an additional factor of $1/(2J+1)$ to calculate the decay width.

\section{Results}
\label{sec:res}

\subsection{Considering only the dominant channel}

First, we show the results of the $P_c$ decay into $J/\psi N$ with a simplification in Eq.~\eqref{eq:AiJ}, i.e.,
we approximate the sum over $X$ for the intermediate states by considering only the channel which has the largest coupling to $P_{ci}$ (as listed in Tables~\ref{tab:pole1} and~\ref{tab:pole2} for Case~1 and Case~2, respectively). Then the decay amplitude is 
\begin{align}
 \mA_{i(J)}=g_{P_{ci},X} \tilde{G}_X g_1h_{X(J)},\label{eq:AiJapp}
\end{align}
with $X=\bar{D}\Sigma_c~(i=1),~\bar{D}\Sigma_c^*~(i=2),~\bar{D}^*\Sigma_c~(i=3,4)$, and $\bar{D}^*\Sigma_c^*~(i=5,6)$.

A few remarks are in order here. In the single-channel case, the effective coupling constant of the $P_{ci}$ state to the constituent channel $X$ is related to the binding energy $E_{Bi}\equiv m_X +M_X - M_{P_{ci}}$, with $m_X$ and $M_X$ being the meson and baryon masses in channel $X$ and $M_{P_{ci}}$ being the mass of $P_{ci}$, as $g_{P_{ci},X}^2\propto \sqrt{E_{B,i}}$~\cite{Weinberg:1965zz} (see, e.g., Sections III.B and VI.B of Ref.~\cite{Guo:2017jvc}). The nonrelativistic loop integral $\tilde{G}_X$ is linearly divergent; working out the regularized integral in Eq.~\eqref{eq:G}, one gets 
\begin{equation}
    \tilde{G}_X\propto -\frac{\Lambda}{\sqrt{\pi}} + \sqrt{2\mu_X E_{B,i}} + \mathcal{O}(\Lambda^{-1}) \,.
    \label{eq:Gexp}
\end{equation}
If we keep only the LO term in the expansion in powers of $\sqrt{2\mu_X E_{B,i}}/\Lambda$, the $\Lambda$-dependence can be absorbed by $g_1$, which needs to scale as $1/\Lambda$, via a multiplicative renormalization. As a result, at LO, the product $G_Xg_1$ is independent of $\Lambda$, and we obtain the following factorization formula,\footnote{This is similar to the  factorization formula for the production of the $X(3872)$ in $B$ decays discussed in Ref.~\cite{Braaten:2005jj}.}
\begin{equation}
    |\mA_{i(J)}|^2 \propto \sqrt{E_{Bi}} h_{X(J)}^2 \,,
    \label{eq:simple}
\end{equation}
where the factor $\propto g_1/\Lambda$ encoding the short-distance physics is not shown, and the factor $\sqrt{E_{Bi}}$ encodes the long-distance physics from the hadronic molecular nature. Its physical meaning is as follows: decreasing the binding energy, the size of the hadronic molecule increases; then its decay by recombining the quark contents in the two constituent hadrons becomes more difficult, and the decay rate decreases with a speed proportional to the square root of the binding energy.

With the above formula, one can easily work out ratios of the partial widths of different $P_c$ states into the $J/\psi N$. However, we notice that different phase space factors should be taken into account for different $P_c$, and there are cases with a binding energy as large as about 20~MeV such that the binding momentum is about 0.2~GeV. Then the higher-order terms in Eq.~\eqref{eq:Gexp} can have sizable contributions. Thus, we use the full expression of Eq.~\eqref{eq:G} and take two values of $\Lambda$, $0.7$~GeV and $1$~GeV, as discussed below that equation, to check the cutoff dependence. 
Defining $r_i$ with $r_i=\Gamma_i/\Gamma_1$ $(i=2,\ldots, 6)$,
we obtain 
\begin{align}
 \begin{split}
  \cs~1:~&r_2=(4.6, 4.6),~r_3=(14.6, 15.3),~r_4=(1.2,1.2),~r_5=(5.4,5.8),~r_6=(10.3,10.9),\\ 
  \cs~2:~&r_2=(4.5, 4.6),~r_3=(6.2, 6.3),~r_4=(1.9, 2.0),~r_5=(1.8, 1.9),~r_6=(7.0, 7.3),
  \end{split}
 \label{eq:risngl}
\end{align}
where the first and second numbers in parentheses are obtained using $\Lambda=0.7$~GeV and $\Lambda=1$~GeV, respectively. One sees that the dependence of the results on the cutoff value is weak.  
The values given above are in line with the simple expectation in Eq.~\eqref{eq:simple}. 
Numerical differences can be traced back to the difference of binding energies and phase space factors for the $P_c$ states as mentioned above.

\subsection{Including all channels}

When all of the coupled channels are included, the qualitative features of the ratios are the same as those in the above single-channel calculation, though the numerical values change to 
\begin{align}
 \begin{split}
  \cs~1:~&r_2=(4.4,4.4),~r_3=(9.3,9.6),~r_4=(1.2,1.2),~r_5=(2.7,2.8),~r_6=(5.1,5.4),\\
  \cs~2:~&r_2=(4.6,4.7),~r_3=(10.1,10.5),~r_4=(1.9,1.9),~r_5=(3.3,3.5),~r_6=(13.9,14.4),
 \end{split}
 \label{ririjfull}
\end{align}
where again the first and second numbers in parentheses are obtained using $\Lambda=0.7$~GeV and $\Lambda=1$~GeV, respectively.

\subsection{Discussions}

Note that $P_{c3}$ $[P_{c4}]$ is assigned as the $P_c(4440)$ $[P_c(4457)]$ in Case~1, 
and $P_{c4}$ $[P_{c3}]$ is assigned as the $P_c(4440)$ $[P_c(4457)]$ in Case~2;
in both cases, $P_{c1}$ refers to the $P_c(4312)$.
From the above numerical results, one finds that the partial widths of $P_{c3}$ and $P_{c4}$, i.e., $P_c(4440)$ and $P_c(4457)$, into the $J/\psi N$ are very different, and at least one of them is much larger than that of the $P_c(4312)$.  
In the measured $J/\psi p$ invariant mass distribution of the $\Lambda_b^0\to K^- J/\psi p$ decay~\cite{Aaij:2019vzc}, there are only three clear peaks corresponding to  the $P_c(4312)$, $P_c(4440)$ and $P_c(4457)$. The ratio of branching fractions $\mathcal{B}(\Lambda_b^0\to K^- P_c^+) \mathcal{B}(P_c^+\to J/\psi p) /\mathcal{B}(\Lambda_b^0\to K^- J/\psi p)$ was measured to be $0.30^{+0.35}_{-0.11}$, $1.11^{+0.40}_{-0.34}$, and $0.53^{+0.22}_{-0.21}$ for $P_c(4312)$, $P_c(4440)$, and $P_c(4457)$, respectively, where the statistical and systematic errors in Ref.~\cite{Aaij:2019vzc} have been added in quadrature. Using the values in Eq.~\eqref{ririjfull}, we obtain the ratios $\frac{\mathcal{B}(\Lambda_b^0\to K^- P_c(4457)^+)}{\Gamma(P_c(4457)^+)}: \frac{\mathcal{B}(\Lambda_b^0\to K^- P_c(4440)^+)}{\Gamma(P_c(4440)^+)}:\frac{\mathcal{B}(\Lambda_b^0\to K^- P_c(4312)^+)}{\Gamma(P_c(4312)^+)}$ as 
\begin{equation}
  \cs~1:~~ 1.5: 0.4 : 1 \,, \qquad
  \cs~2:~~ 0.2:1.9 :1 \,,
 \label{eq:lampc}
\end{equation}
where only the central values are shown. One sees that the ratio of ${\mathcal{B}(\Lambda_b^0\to K^- P_c^+)}/{\Gamma(P_c^+)}$ can differ by one order of magnitude in Case 2.

\begin{figure}[t]
 \centering
 \includegraphics[height=3cm]{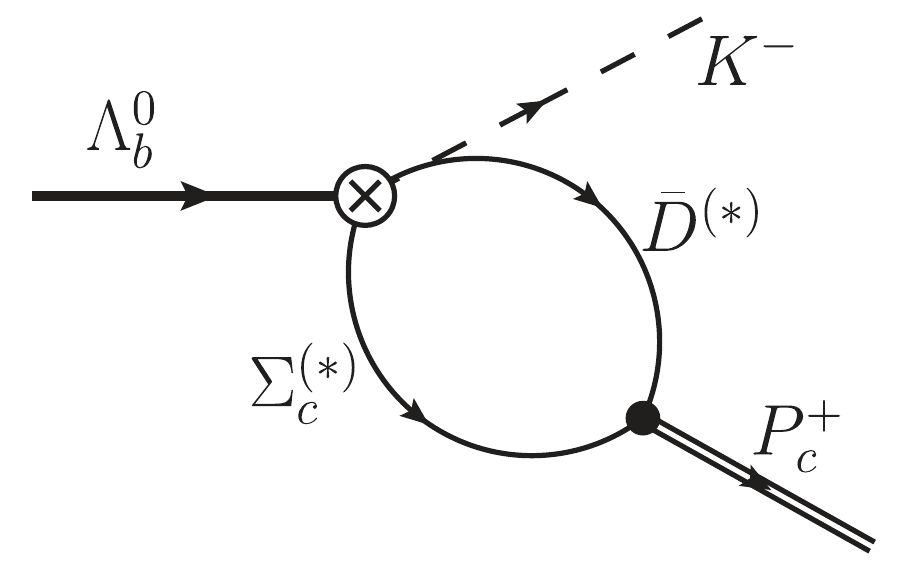} \hspace{1cm}
 \includegraphics[height=3cm]{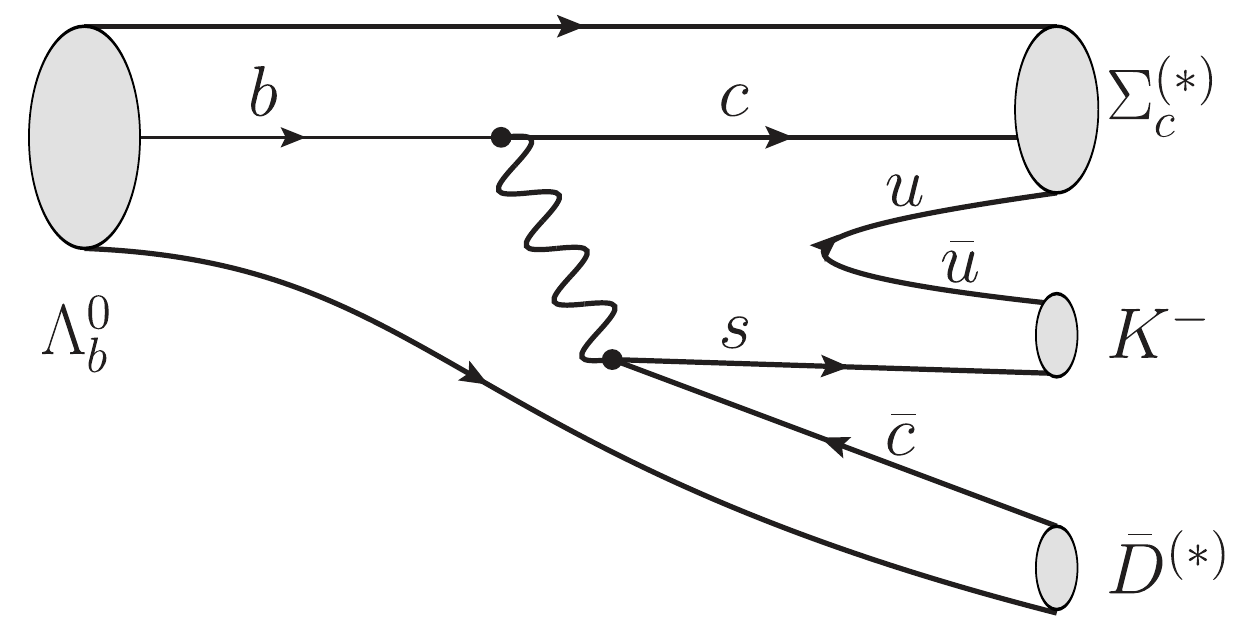}
 \caption{Left: Production of the $P_c$ from the $\Lambda_b$ decay in the hadronic molecular picture through intermediate $\bar{D}^{(*)}\Sigma_c^{(*)}$ states. Right: A possible quark-level diagram for the weak decay $\Lambda_b^0\to K^-\Sigma_c^{(*)}\bar D^{(*)}$.}
 \label{fig2}
\end{figure}

The production mechanism of the $P_c$ states from $\Lambda_b^0$ decays in the hadronic molecular model is shown in Fig.~\ref{fig2}. Using the same arguments leading to Eq.~\eqref{eq:simple}, one gets the factorization formula for the production rate as the product of a short-distance part and a long-distance part. The long-distance part is proportional to the square root of the binding energy, as is that for the decay; see Eq.~\eqref{eq:simple}. However, the short-distance part  differs for different $P_c$ states even though some of them couple dominantly to the same $\Sigma_c^{(*)}\bar D^{(*)}$ pair, as can be seen from the fact that different partial waves are involved in the decays of the $\Lambda_b^0$ into $K^-$ and $P_c^+$ with different spins. This makes it difficult to relate the productions of different $P_c$ states to each another.
For a model calculation of the $\Lambda_b$ decays into the three observed $P_c$ states, see Ref.~\cite{Wu:2019rog}.

Moreover, one finds that the partial widths of the $P_{c3}$, $P_{c5}$ and $P_{c6}$ are all much larger than that of $P_{c1}$, i.e., $P_c(4312)$.
However, no visible $P_c$ peaks around 4.50 to 4.52~GeV (the mass region of $P_{c5,c6}$ in both Case~1 and Case~2) can be seen in the $J/\psi p$ invariant mass distribution. This indicates that either the $P_{c5}$ and $P_{c6}$ states are much more difficult to produce than the $P_c(4312)$ or there are other mechanisms for producing the observed three $P_c$ states.
One possibility is that the observed peaking structures are a result of an intricate interplay between the  $\bar D^{(*)}\Sigma_c$ hadronic molecules and the triangle singularities discussed in Refs.~\cite{Guo:2015umn,Liu:2015fea,Guo:2016bkl,Bayar:2016ftu} (see also Appendix of Ref.~\cite{Aaij:2019vzc}), with the latter providing an enhancement at around 4.45~GeV.

\subsection{Decays into \texorpdfstring{$\bm{\eta_c N}$}{etac N}}

The partial decay widths of $P_{ci}$ into $\eta_cN$ $(i=1,3,5)$ normalized to the $P_{c1}\rightarrow J/\psi N$ partial width can also be obtained in the same way. Letting
$\tilde{r}_i=\tilde{\Gamma}_i/\Gamma_1$ with $\Gamma_i$ and $\tilde{\Gamma}_i$ in Eqs.~\eqref{eq:Gami} and \eqref{eq:Gamtili}, we get 
\begin{align}
 {\rm Case}~1:&\quad \tilde{r}_1=(2.9,2.9),\quad \tilde{r}_3=(0.4,0.4), \quad \tilde{r}_5=(9.8,10.3),\\
 {\rm Case}~2:&\quad \tilde{r}_1=(4.0,4.0), \quad \tilde{r}_3=(2.4,2.5), \quad \tilde{r}_5=(10.2,10.7).
\end{align}
One finds that the partial width of the $P_c(4312)\to \eta_c N$ is larger than that of the $J/\psi N$ mode (see also Ref.~\cite{Voloshin:2019aut}).\footnote{In the pioneering works predicting the existence of hidden-charm pentaquarks~\cite{Wu:2010jy,Wu:2010vk}, the authors already noticed that the $\Sigma_c\bar D$ hadronic molecule decays more easily into the $\eta_c N$ than into the $J/\psi N$. Since not all of the HQSS related channels were considered therein, the predicted ratio differs a lot from our result.}
In both cases, we expect significant peaks to appear around $4.3~\gev$ from $P_c(4312)$ and around $4.5~\gev$ from a $\bar{D}^*\Sigma_c^*$ molecule
if the background is of the same order as in the $J/\psi N$ case and the productions are similar.
The ratio $\tilde{r}_3$ is smaller (larger) than 1 in Case~1 (Case~2) (recall that $P_{c3}$ refers to the $P_c(4440)$ decay in Case~1, and to the $P_c(4457)$ decay in Case~2). 
Thus, a search of hidden-charm pentaquarks in  the $\eta_cN$ channel can shed light on the origin of the $P_c$ states.

\section{Summary}
\label{sec:sum}
We investigate in this paper the decays of the $\bar{D}^{(*)}\Sigma_c^{(*)}$ molecular $P_c$ states into the $J/\psi N$ and $\eta_cN$ final states with a setup respecting HQSS. We use the coupled-channel ($\bar{D}^{(*)}\Sigma_c^{(*)}$) Lippmann-Schwinger equation, and the $P_c$ states are obtained as poles of the $T$-matrix. Following Refs.~\cite{Liu:2019tjn,Xiao:2019aya}, model parameters are fixed to reproduce the peak positions of the $P_c(4440)$ and the $P_c(4457)$, and five additional states with binding energies ranging from a few to about 20~MeV are obtained as a consequence of HQSS~\cite{Xiao:2013yca,Liu:2019tjn,Xiao:2019aya}. 
{Some of the seven poles may move into a ``wrong'' Riemann sheet within a $25\%$ uncertainty of the low-energy constants accounting for the HQSS breaking effects.
Here we stress that the poles of $\bar{D}\Sigma_c$ and $\bar{D}\Sigma_c^*$ molecules always exist in the correct Riemann sheet,
and one or two of three poles close to the $\bar{D}^*\Sigma_c^*$ threshold remain as well.}
The lowest pole has a mass consistent with that of the $P_c(4312)$, and it couples dominantly to the $\bar D\Sigma_c$ with $J^P=1/2^-$. 
The $P_c(4440)$ and $P_c(4457)$ couple dominantly to the $\bar D^*\Sigma_c$, and their quantum numbers are $1/2^-$ and $3/2^-$. Two possible assignments of $P_c(4440)$ and $P_c(4457)$ are considered as in Ref.\cite{Liu:2019tjn}: 
in one case the spins of $P_c(4440)$ and $P_c(4457)$ are  $1/2$ and $3/2$, respectively, and in the other case the ordering is reversed. Among all seven $P_c$ states, six (with $J^P=1/2^-,3/2^-$) can decay into the $J/\psi N$ in an $S$-wave, and three (with $J^P=1/2^-$) can decay into the $\eta_c N$ in an $S$-wave. HQSS allows us to predict parameter-free ratios of the partial widths of these decays. It is found that  five $P_c$ states with $J^P=1/2^-,3/2^-$ decay into the $J/\psi N$ more easily than the $P_c(4312)$, and the $P_c(4312)$ decays into the $\eta_c N$ with a partial width three times that of the $J/\psi N$ mode. We find that the partial widths into the $J/\psi N$ for the $J^P=1/2^-$ $\bar D\Sigma_c^*$ molecule with a mass around 4.37~GeV and the $J^P=1/2^-$ and $3/2^-$ $\bar D^*\Sigma_c^*$ molecules with masses in the range of 4.50 to 4.52~GeV are all larger than that for the $P_c(4312)$. The nonobservation of any of them could be because they have smaller production rates from the $\Lambda_b$ decays, or because the observed peaks receive contributions from other mechanisms such as triangle singularities in addition to the hidden-charm pentaquarks. In order to reveal the nature of the observed pentaquark candidates, more measurements and a detailed amplitude analysis considering both resonances and kinematical singularities are called for. The results in this paper provide useful input into search for more $P_c$ states in the $J/\psi p$ and $\eta_c p$ final states.

\medskip

\begin{acknowledgements}
This work is supported in part by the National Natural Science Foundation of China (NSFC) and  the Deutsche Forschungsgemeinschaft (DFG) through the funds provided to the Sino-German Collaborative Research Center ``Symmetries and the Emergence of Structure in QCD"  (NSFC Grant No. 11621131001 and DFG Grant No. TRR110), by the NSFC under Grants No. 11847612 and No. 11835015, by the Chinese Academy of Sciences (CAS) under Grants No. QYZDB-SSW-SYS013 and No. XDPB09, and by
the CAS Center for Excellence in Particle Physics (CCEPP). S.S. is also supported by {2019 International Postdoctoral Exchange Program, and} by the CAS President’s International Fellowship Initiative (PIFI) under Grant No.~2019PM0108.
\end{acknowledgements}

\medskip

\appendix
\section{\texorpdfstring{$\bm{\bar{D}^{(*)}\Sigma_c^{(*)}}$}{Dbar Sigmac} interaction}
\label{app:interaction}

The construction of interaction vertices by rearranging the heavy quark and light quark spins respecting HQSS using the $9j$ symbol is used in, e.g., Refs.~\cite{Xiao:2013yca,Guo:2017jvc,Lu:2017dvm}.
In the $\bar{D}^{(*)}$ meson, the heavy quark component has spin $s_H^{\bar{D}^{(*)}}=1/2$, and the light quark component has spin $s_L^{\bar{D}^{(*)}}=1/2$;
in the $\Sigma_c^{(*)}$ baryon, the heavy quark component has spin $s_H^{\Sigma_c^{(*)}}=1/2$, and the light quark component has spin $s_L^{\Sigma_c^{(*)}}=1$.
The $\bar{D}^{(*)}\Sigma_c^{(*)}$ system with spin $J$ can be specified with the spins of its constituents as
\begin{align}
 \left|\bar{D}^{(*)}\Sigma_c^{(*)}\right>_J=\left|s_H^{\bar{D}^{(*)}},s_L^{\bar{D}^{(*)}},j^{\bar{D}^{(*)}};s_H^{\Sigma_c^{(*)}},s_L^{\Sigma_c^{(*)}},j^{\Sigma_c^{(*)}};J\right>,
\end{align}
where $j^{\bar{D}^{(*)}}$ and $j^{\Sigma_c^{(*)}}$ are the spins of the $\bar{D}^{(*)}$ and the $\Sigma_c^{(*)}$, respectively.
Using the $9j$ symbol, this state can be rewritten with a linear combination of the eigenstates of the spin of $c\bar{c}$, $s_H$, and that of the light degrees of freedom, $s_L$;
\begin{align}
 \begin{split}
  &\left|s_H^{\bar{D}^{(*)}},s_L^{\bar{D}^{(*)}},j^{\bar{D}^{(*)}};s_H^{\Sigma_c^{(*)}},s_L^{\Sigma_c^{(*)}},j^{\Sigma_c^{(*)}};J\right>\\
  &=\sum_{s_L,s_H}\sqrt{(2s_L+1)(2s_H+1)(2j^{\bar{D}^{(*)}}+1)(2j^{\Sigma_c^{(*)}}+1)}\\
  &\times\left\{ 
  \begin{array}{ccc}
   s_L^{\bar{D}^{(*)}} &s_L^{\Sigma_c^{(*)}}& s_L\\
   s_H^{\bar{D}^{(*)}} &s_H^{\Sigma_c^{(*)}}& s_H\\
   j^{\bar{D}^{(*)}}&j^{\Sigma_c^{(*)}} &J 
  \end{array} 
  \right\}
  \left|s_L^{\bar{D}^{(*)}},s_L^{\Sigma_c^{(*)}},s_L;s_H^{\bar{D}^{(*)}},s_H^{\Sigma_c^{(*)}},s_H;J\right>,
 \end{split}
\end{align} 
where $\left\{...\right\}$ denotes Wigner's $9j$ symbol.
Then, the $\bar{D}^{(*)}\Sigma_c^{(*)}$ states with spin $J=1/2$, $3/2$, and $5/2$ are expressed in terms of the $s_H$ and $s_L$ eigenstates as follows:
\begin{align}
 \left|\bar{D}\Sigma_c\right>_{1/2}=&\frac{1}{2}\left|1/2_L,0_H;1/2_J\right>-\frac{1}{2\sqrt{3}}\left|1/2_L,1_H;1/2_J\right>+\sqrt{\frac{2}{3}}\left|3/2_L,1_H;1/2_J\right>,\\
 \left|\bar{D}\Sigma_c^*\right>_{3/2}=&\frac{1}{2}\left|3/2_L,0_H;3/2_J\right>-\frac{1}{\sqrt{3}}\left|1/2_L,1_H;3/2_J\right>+\frac{1}{2}\sqrt{\frac{5}{3}}\left|3/2_L,1_H;3/2_J\right>,\\
 \left|\bar{D}^*\Sigma_c\right>_{1/2}=&-\frac{1}{2\sqrt{3}}\left|1/2_L,0_H;1/2_J\right>+\frac{5}{6}\left|1/2_L,1_H;1/2_J\right>+\frac{\sqrt{2}}{3}\left|3/2_L,1_H;1/2_J\right>,\\
 \left|\bar{D}^*\Sigma_c\right>_{3/2}=&-\frac{1}{\sqrt{3}}\left|3/2_L,0_H;3/2_J\right>+\frac{1}{3}\left|1/2_L,1_H;3/2_J\right>+\frac{\sqrt{5}}{3}\left|3/2_L,1_H;3/2_J\right>,\\
 \left|\bar{D}^*\Sigma_c^*\right>_{1/2}=&\sqrt{\frac{2}{3}}\left|1/2_L,0_H;1/2_J\right>+\frac{\sqrt{2}}{3}\left|1/2_L,1_H;1/2_J\right>-\frac{1}{3}\left|3/2_L,1_H;1/2_J\right>,\\
 \left|\bar{D}^*\Sigma_c^*\right>_{3/2}=&\frac{1}{2}\sqrt{\frac{5}{3}}\left|3/2_L,0_H;3/2_J\right>+\frac{\sqrt{5}}{3}\left|1/2_L,1_H;3/2_J\right>+\frac{1}{6}\left|3/2_L,1_H;3/2_J\right>,\\
 \left|\bar{D}^*\Sigma_c^*\right>_{5/2}=&\left|3/2_L,1_H;5/2_J\right>.
\end{align}
On the right-hand side of these equations, the trivial arguments $s_{L(H)}^{\bar{D}^{(*)}(\Sigma_c^{(*)})}$ are suppressed, i.e., only $s_L$, $s_H$, and $J$ are shown explicitly.

In the heavy quark limit, the spins of heavy quarks decouple from the dynamics, and the interaction only depends on the spin of light degrees of freedom $s_L$ (both $s_L$ and $s_H$ are conserved).
We can write the matrix element $\left<s_L,s_H;J\right|V_{\rm int}\left|s_L',s_H';J'\right>=C_{(2s_L+1)/2}\delta_{J,J'}\delta_{s_H,s'_H}\delta_{s_L,s'_{L}}$ (now we suppress the isospin index because we consider the $I=1/2$ case only).
With the substitution of $C_1=C_a-2C_b$ and $C_2=C_a+C_b$, one can obtain the transition amplitude from channel $X$ to $X'$ $(X,X'=\bar{D}^{(*)}\Sigma_c^{(*)})$, $v_{X,X'(J)}$ in Eq.~\eqref{eq:v}, as summarized in 
Table~\ref{table:CG}.

Here, we note that the meson fields are normalized in the nonrelativistic way, 
and the interaction $v$ in Eq.~\eqref{eq:SLeq} is $\sqrt{(2m_X)(2m_{X'})}v_{X,X'(J)}$ with $v_{X,X'(J)}$ in Eq.~\eqref{eq:v} ($m_X$ is the meson mass in channel $X$).
Then, $C_{a,b}$ have a dimension mass$^{-2}$ and $v$ has mass$^{-1}$ in our calculation.

One can also start from the effective Lagrangian given in Ref.~\cite{Liu:2018zzu},
\begin{align}
 \mL_{\bar{D}^{(*)}\Sigma_c^{(*)},\bar{D}^{(*)}\Sigma_c^{(*)}}=-C_a\vec{S}_c^\dagger\cdot\vec{S}_c\tr[\bar{H}_c^\dagger\bar{H}_c]-C_bi\epsilon_{jik}(S_c^\dagger)_j(S_c)_k\tr[\bar{H}_c^\dagger\sigma_i\bar{H}_c]\, ,\label{eq:LHQSS}
\end{align}
where $\sigma_i$ $(i=1,2,3)$ are the Pauli matrices, and $\vec{S}_c$ and $\bar{H}_c$ are the  heavy quark spin doublets of $(\Sigma_c,\Sigma_c^*)$ and $(\bar{D},\bar{D}^*)$ in the two-component notation~\cite{Hu:2005gf}
(see, e.g., Refs.~\cite{Falk:1992cx,Cho:1992cf,Valderrama:2012jv} for the four-component notation),
\begin{align}
 \vec{S}_c=&\frac{1}{\sqrt{3}}\vec{\sigma}\,\Sigma_c+\vec{\Sigma}^*_c\,,\label{eq:Scvec}\\
 \bar{H}_c=&\frac{1}{\sqrt{2}}\left(-\bar{D}+\vec{\sigma}\cdot\vec{\bar{D}}^*\right)\,.
 \label{eq:Hcbar}
\end{align}
This Lagrangian, Eq.~\eqref{eq:LHQSS}, gives the $\bar D^{(*)}\Sigma_c^{(*)}$ $S$-wave interaction which is the leading order of the momentum expansion.
The $C_a$ and $C_b$ terms come from the vector and axial-vector currents.

To see the relationship to the coefficient obtained with the $9j$ symbol, we perform a spin projection and average over polarizations.
Writing the amplitude of the $X\rightarrow X'$ transition  $(X^{(\prime)}=\bar{D}\Sigma_c,\bar{D}\Sigma_c^*,\bar{D}^*\Sigma_c,\bar{D}^*\Sigma_c^*)$ given by the Lagrangian Eq.~\eqref{eq:LHQSS} as $t_{X',X}^{(\lambda's',\lambda s)}$ [$s(s')$ and $\lambda(\lambda')$ denote the third components of the spins of the baryon and meson in the initial (final) state, respectively], we give the projection of the amplitude on spin $J$ as
\begin{align}
t_{X',X}^{(Jj)}=&\sum_{\lambda,\lambda',s,s'}\mC(j_{X'}^{\bar{D}^{(*)}},j_{X'}^{\Sigma^{(*)}},J;\lambda',s',j)
\mC(j_{X}^{\bar{D}^{(*)}},j_X^{\Sigma^{(*)}},J;\lambda,s,j)t_{X',X}^{(\lambda's',\lambda s)},\label{eq:spinprj}
\end{align}
where $j$ is the third component of spin $J$, $\mC$ is the Clebsch-Gordan coefficient, and $j_{X^{(\prime)}}^{\bar{D}^{(*)}(\Sigma_c^{(*)})}$ is the spin of $\bar{D}^{(*)}(\Sigma_c^{(*)})$ in the channel $X^{(\prime)}$.
The polarization average of the amplitude is given by
\begin{align}
\bar{t}_{X',X}^{(J)}=&\frac{1}{2J+1}\sum_j t_{X',X}^{(Jj)}.\label{eq:spinav}
\end{align}
This spin averaged amplitude provides the same result as that obtained using the $9j$ symbol given by Eq.~\eqref{eq:v}.

For the transition of $\bar{D}^{(*)}\Sigma_c^{(*)}$ into $\eta_c N$ or $J/\psi N$ we provide the decomposition of the $\eta_c N$ and $J/\psi N$:
\begin{align}
 \left|\eta_c N\right>_{1/2}=&\left|1/2_L,0_H;1/2_J\right>,\\
 \left|J/\psi N\right>_{1/2}=&\left|1/2_L,1_H;1/2_J\right>,\\
 \left|J/\psi N\right>_{3/2}=&\left|1/2_L,1_H;3/2_J\right>.
\end{align}
The matrix element $\left<1/2_L,0_H(1_H);1/2_J\right|V_{\rm int}\left|1/2_L,0_H(1_H);1/2_J\right>$ is denoted by the parameter $g_1$ in Eqs.~\eqref{eq:tXJN} and \eqref{eq:tXecN},
which is independent of $s_H$.

The coefficients of the $X\rightarrow J/\psi N$ and $\eta_c N$ transitions, $h_{X(J)}$ and $\tilde{h}_{X(J)}$ in Eqs.~\eqref{eq:tXJN} and \eqref{eq:htilX}, can be obtained by using the following effective Lagrangian respecting HQSS with projection on spin and average over polarizations in the same manner as in Eqs.~\eqref{eq:spinprj} and \eqref{eq:spinav}:
\begin{align}
 \mL_{\bar{D}^{(*)}\Sigma_c^{(*)},J/\psi N}
 =&\frac{g_1}{\sqrt{6}}N^\dagger\sigma_i\bar{H}_cJ^\dagger(S_c)_i,
\end{align}
where $N$ denotes the nucleon field and $J=-\eta_c+\vec{\sigma}\cdot\vec{\psi}$ is a doublet composed of $\eta_c$ and $J/\psi$~\cite{Casalbuoni:1996pg,Guo:2010ak}.

\bigskip

\bibliography{biblio}

\begin{thebibliography}{80}%
\makeatletter
\providecommand \@ifxundefined [1]{%
 \@ifx{#1\undefined}
}%
\providecommand \@ifnum [1]{%
 \ifnum #1\expandafter \@firstoftwo
 \else \expandafter \@secondoftwo
 \fi
}%
\providecommand \@ifx [1]{%
 \ifx #1\expandafter \@firstoftwo
 \else \expandafter \@secondoftwo
 \fi
}%
\providecommand \natexlab [1]{#1}%
\providecommand \enquote  [1]{``#1''}%
\providecommand \bibnamefont  [1]{#1}%
\providecommand \bibfnamefont [1]{#1}%
\providecommand \citenamefont [1]{#1}%
\providecommand \href@noop [0]{\@secondoftwo}%
\providecommand \href [0]{\begingroup \@sanitize@url \@href}%
\providecommand \@href[1]{\@@startlink{#1}\@@href}%
\providecommand \@@href[1]{\endgroup#1\@@endlink}%
\providecommand \@sanitize@url [0]{\catcode `\\12\catcode `\$12\catcode
  `\&12\catcode `\#12\catcode `\^12\catcode `\_12\catcode `\%12\relax}%
\providecommand \@@startlink[1]{}%
\providecommand \@@endlink[0]{}%
\providecommand \url  [0]{\begingroup\@sanitize@url \@url }%
\providecommand \@url [1]{\endgroup\@href {#1}{\urlprefix }}%
\providecommand \urlprefix  [0]{URL }%
\providecommand \Eprint [0]{\href }%
\providecommand \doibase [0]{http://dx.doi.org/}%
\providecommand \selectlanguage [0]{\@gobble}%
\providecommand \bibinfo  [0]{\@secondoftwo}%
\providecommand \bibfield  [0]{\@secondoftwo}%
\providecommand \translation [1]{[#1]}%
\providecommand \BibitemOpen [0]{}%
\providecommand \bibitemStop [0]{}%
\providecommand \bibitemNoStop [0]{.\EOS\space}%
\providecommand \EOS [0]{\spacefactor3000\relax}%
\providecommand \BibitemShut  [1]{\csname bibitem#1\endcsname}%
\let\auto@bib@innerbib\@empty
\bibitem [{\citenamefont {Aaij}\ \emph {et~al.}(2015)\citenamefont {Aaij} \emph
  {et~al.}}]{Aaij:2015tga}%
  \BibitemOpen
  \bibfield  {author} {\bibinfo {author} {\bibfnamefont {R.}~\bibnamefont
  {Aaij}} \emph {et~al.} (\bibinfo {collaboration} {LHCb}),\ }\href {\doibase
  10.1103/PhysRevLett.115.072001} {\bibfield  {journal} {\bibinfo  {journal}
  {Phys. Rev. Lett.}\ }\textbf {\bibinfo {volume} {115}},\ \bibinfo {pages}
  {072001} (\bibinfo {year} {2015})},\ \Eprint
  {http://arxiv.org/abs/1507.03414} {arXiv:1507.03414 [hep-ex]} \BibitemShut
  {NoStop}%
\bibitem [{\citenamefont {Aaij}\ \emph {et~al.}(2019)\citenamefont {Aaij} \emph
  {et~al.}}]{Aaij:2019vzc}%
  \BibitemOpen
  \bibfield  {author} {\bibinfo {author} {\bibfnamefont {R.}~\bibnamefont
  {Aaij}} \emph {et~al.} (\bibinfo {collaboration} {LHCb}),\ }\href {\doibase
  10.1103/PhysRevLett.122.222001} {\bibfield  {journal} {\bibinfo  {journal}
  {Phys. Rev. Lett.}\ }\textbf {\bibinfo {volume} {122}},\ \bibinfo {pages}
  {222001} (\bibinfo {year} {2019})},\ \Eprint
  {http://arxiv.org/abs/1904.03947} {arXiv:1904.03947 [hep-ex]} \BibitemShut
  {NoStop}%
\bibitem [{\citenamefont {Guo}\ \emph {et~al.}(2018)\citenamefont {Guo},
  \citenamefont {Hanhart}, \citenamefont {Meißner}, \citenamefont {Wang},
  \citenamefont {Zhao},\ and\ \citenamefont {Zou}}]{Guo:2017jvc}%
  \BibitemOpen
  \bibfield  {author} {\bibinfo {author} {\bibfnamefont {F.-K.}\ \bibnamefont
  {Guo}}, \bibinfo {author} {\bibfnamefont {C.}~\bibnamefont {Hanhart}},
  \bibinfo {author} {\bibfnamefont {U.-G.}\ \bibnamefont {Meißner}}, \bibinfo
  {author} {\bibfnamefont {Q.}~\bibnamefont {Wang}}, \bibinfo {author}
  {\bibfnamefont {Q.}~\bibnamefont {Zhao}}, \ and\ \bibinfo {author}
  {\bibfnamefont {B.-S.}\ \bibnamefont {Zou}},\ }\href {\doibase
  10.1103/RevModPhys.90.015004} {\bibfield  {journal} {\bibinfo  {journal}
  {Rev. Mod. Phys.}\ }\textbf {\bibinfo {volume} {90}},\ \bibinfo {pages}
  {015004} (\bibinfo {year} {2018})},\ \Eprint
  {http://arxiv.org/abs/1705.00141} {arXiv:1705.00141 [hep-ph]} \BibitemShut
  {NoStop}%
\bibitem [{\citenamefont {Dalitz}\ \emph {et~al.}(1967)\citenamefont {Dalitz},
  \citenamefont {Wong},\ and\ \citenamefont {Rajasekaran}}]{Dalitz:1967fp}%
  \BibitemOpen
  \bibfield  {author} {\bibinfo {author} {\bibfnamefont {R.~H.}\ \bibnamefont
  {Dalitz}}, \bibinfo {author} {\bibfnamefont {T.~C.}\ \bibnamefont {Wong}}, \
  and\ \bibinfo {author} {\bibfnamefont {G.}~\bibnamefont {Rajasekaran}},\
  }\href {\doibase 10.1103/PhysRev.153.1617} {\bibfield  {journal} {\bibinfo
  {journal} {Phys. Rev.}\ }\textbf {\bibinfo {volume} {153}},\ \bibinfo {pages}
  {1617} (\bibinfo {year} {1967})}\BibitemShut {NoStop}%
\bibitem [{\citenamefont {Kaiser}\ \emph {et~al.}(1995)\citenamefont {Kaiser},
  \citenamefont {Siegel},\ and\ \citenamefont {Weise}}]{Kaiser:1995eg}%
  \BibitemOpen
  \bibfield  {author} {\bibinfo {author} {\bibfnamefont {N.}~\bibnamefont
  {Kaiser}}, \bibinfo {author} {\bibfnamefont {P.~B.}\ \bibnamefont {Siegel}},
  \ and\ \bibinfo {author} {\bibfnamefont {W.}~\bibnamefont {Weise}},\ }\href
  {\doibase 10.1016/0375-9474(95)00362-5} {\bibfield  {journal} {\bibinfo
  {journal} {Nucl. Phys.}\ }\textbf {\bibinfo {volume} {A594}},\ \bibinfo
  {pages} {325} (\bibinfo {year} {1995})},\ \Eprint
  {http://arxiv.org/abs/nucl-th/9505043} {arXiv:nucl-th/9505043 [nucl-th]}
  \BibitemShut {NoStop}%
\bibitem [{\citenamefont {Oset}\ and\ \citenamefont
  {Ramos}(1998)}]{Oset:1997it}%
  \BibitemOpen
  \bibfield  {author} {\bibinfo {author} {\bibfnamefont {E.}~\bibnamefont
  {Oset}}\ and\ \bibinfo {author} {\bibfnamefont {A.}~\bibnamefont {Ramos}},\
  }\href {\doibase 10.1016/S0375-9474(98)00170-5} {\bibfield  {journal}
  {\bibinfo  {journal} {Nucl. Phys.}\ }\textbf {\bibinfo {volume} {A635}},\
  \bibinfo {pages} {99} (\bibinfo {year} {1998})},\ \Eprint
  {http://arxiv.org/abs/nucl-th/9711022} {arXiv:nucl-th/9711022 [nucl-th]}
  \BibitemShut {NoStop}%
\bibitem [{\citenamefont {Oller}\ and\ \citenamefont
  {Mei{\ss}ner}(2001)}]{Oller:2000fj}%
  \BibitemOpen
  \bibfield  {author} {\bibinfo {author} {\bibfnamefont {J.~A.}\ \bibnamefont
  {Oller}}\ and\ \bibinfo {author} {\bibfnamefont {U.-G.}\ \bibnamefont
  {Mei{\ss}ner}},\ }\href {\doibase 10.1016/S0370-2693(01)00078-8} {\bibfield
  {journal} {\bibinfo  {journal} {Phys. Lett.}\ }\textbf {\bibinfo {volume}
  {B500}},\ \bibinfo {pages} {263} (\bibinfo {year} {2001})},\ \Eprint
  {http://arxiv.org/abs/hep-ph/0011146} {arXiv:hep-ph/0011146 [hep-ph]}
  \BibitemShut {NoStop}%
\bibitem [{\citenamefont {Hyodo}\ and\ \citenamefont
  {Jido}(2012)}]{Hyodo:2011ur}%
  \BibitemOpen
  \bibfield  {author} {\bibinfo {author} {\bibfnamefont {T.}~\bibnamefont
  {Hyodo}}\ and\ \bibinfo {author} {\bibfnamefont {D.}~\bibnamefont {Jido}},\
  }\href {\doibase 10.1016/j.ppnp.2011.07.002} {\bibfield  {journal} {\bibinfo
  {journal} {Prog. Part. Nucl. Phys.}\ }\textbf {\bibinfo {volume} {67}},\
  \bibinfo {pages} {55} (\bibinfo {year} {2012})},\ \Eprint
  {http://arxiv.org/abs/1104.4474} {arXiv:1104.4474 [nucl-th]} \BibitemShut
  {NoStop}%
\bibitem [{\citenamefont {Kamiya}\ \emph {et~al.}(2016)\citenamefont {Kamiya},
  \citenamefont {Miyahara}, \citenamefont {Ohnishi}, \citenamefont {Ikeda},
  \citenamefont {Hyodo}, \citenamefont {Oset},\ and\ \citenamefont
  {Weise}}]{Kamiya:2016jqc}%
  \BibitemOpen
  \bibfield  {author} {\bibinfo {author} {\bibfnamefont {Y.}~\bibnamefont
  {Kamiya}}, \bibinfo {author} {\bibfnamefont {K.}~\bibnamefont {Miyahara}},
  \bibinfo {author} {\bibfnamefont {S.}~\bibnamefont {Ohnishi}}, \bibinfo
  {author} {\bibfnamefont {Y.}~\bibnamefont {Ikeda}}, \bibinfo {author}
  {\bibfnamefont {T.}~\bibnamefont {Hyodo}}, \bibinfo {author} {\bibfnamefont
  {E.}~\bibnamefont {Oset}}, \ and\ \bibinfo {author} {\bibfnamefont
  {W.}~\bibnamefont {Weise}},\ }\href {\doibase
  10.1016/j.nuclphysa.2016.04.013} {\bibfield  {journal} {\bibinfo  {journal}
  {Nucl. Phys.}\ }\textbf {\bibinfo {volume} {A954}},\ \bibinfo {pages} {41}
  (\bibinfo {year} {2016})},\ \Eprint {http://arxiv.org/abs/1602.08852}
  {arXiv:1602.08852 [hep-ph]} \BibitemShut {NoStop}%
\bibitem [{\citenamefont {Wu}\ \emph {et~al.}(2010)\citenamefont {Wu},
  \citenamefont {Molina}, \citenamefont {Oset},\ and\ \citenamefont
  {Zou}}]{Wu:2010jy}%
  \BibitemOpen
  \bibfield  {author} {\bibinfo {author} {\bibfnamefont {J.-J.}\ \bibnamefont
  {Wu}}, \bibinfo {author} {\bibfnamefont {R.}~\bibnamefont {Molina}}, \bibinfo
  {author} {\bibfnamefont {E.}~\bibnamefont {Oset}}, \ and\ \bibinfo {author}
  {\bibfnamefont {B.~S.}\ \bibnamefont {Zou}},\ }\href {\doibase
  10.1103/PhysRevLett.105.232001} {\bibfield  {journal} {\bibinfo  {journal}
  {Phys. Rev. Lett.}\ }\textbf {\bibinfo {volume} {105}},\ \bibinfo {pages}
  {232001} (\bibinfo {year} {2010})},\ \Eprint {http://arxiv.org/abs/1007.0573}
  {arXiv:1007.0573 [nucl-th]} \BibitemShut {NoStop}%
\bibitem [{\citenamefont {Wu}\ \emph {et~al.}(2011)\citenamefont {Wu},
  \citenamefont {Molina}, \citenamefont {Oset},\ and\ \citenamefont
  {Zou}}]{Wu:2010vk}%
  \BibitemOpen
  \bibfield  {author} {\bibinfo {author} {\bibfnamefont {J.-J.}\ \bibnamefont
  {Wu}}, \bibinfo {author} {\bibfnamefont {R.}~\bibnamefont {Molina}}, \bibinfo
  {author} {\bibfnamefont {E.}~\bibnamefont {Oset}}, \ and\ \bibinfo {author}
  {\bibfnamefont {B.~S.}\ \bibnamefont {Zou}},\ }\href {\doibase
  10.1103/PhysRevC.84.015202} {\bibfield  {journal} {\bibinfo  {journal} {Phys.
  Rev.}\ }\textbf {\bibinfo {volume} {C84}},\ \bibinfo {pages} {015202}
  (\bibinfo {year} {2011})},\ \Eprint {http://arxiv.org/abs/1011.2399}
  {arXiv:1011.2399 [nucl-th]} \BibitemShut {NoStop}%
\bibitem [{\citenamefont {Wang}\ \emph {et~al.}(2011)\citenamefont {Wang},
  \citenamefont {Huang}, \citenamefont {Zhang},\ and\ \citenamefont
  {Zou}}]{Wang:2011rga}%
  \BibitemOpen
  \bibfield  {author} {\bibinfo {author} {\bibfnamefont {W.~L.}\ \bibnamefont
  {Wang}}, \bibinfo {author} {\bibfnamefont {F.}~\bibnamefont {Huang}},
  \bibinfo {author} {\bibfnamefont {Z.~Y.}\ \bibnamefont {Zhang}}, \ and\
  \bibinfo {author} {\bibfnamefont {B.~S.}\ \bibnamefont {Zou}},\ }\href
  {\doibase 10.1103/PhysRevC.84.015203} {\bibfield  {journal} {\bibinfo
  {journal} {Phys. Rev.}\ }\textbf {\bibinfo {volume} {C84}},\ \bibinfo {pages}
  {015203} (\bibinfo {year} {2011})},\ \Eprint {http://arxiv.org/abs/1101.0453}
  {arXiv:1101.0453 [nucl-th]} \BibitemShut {NoStop}%
\bibitem [{\citenamefont {Yang}\ \emph {et~al.}(2012)\citenamefont {Yang},
  \citenamefont {Sun}, \citenamefont {He}, \citenamefont {Liu},\ and\
  \citenamefont {Zhu}}]{Yang:2011wz}%
  \BibitemOpen
  \bibfield  {author} {\bibinfo {author} {\bibfnamefont {Z.-C.}\ \bibnamefont
  {Yang}}, \bibinfo {author} {\bibfnamefont {Z.-F.}\ \bibnamefont {Sun}},
  \bibinfo {author} {\bibfnamefont {J.}~\bibnamefont {He}}, \bibinfo {author}
  {\bibfnamefont {X.}~\bibnamefont {Liu}}, \ and\ \bibinfo {author}
  {\bibfnamefont {S.-L.}\ \bibnamefont {Zhu}},\ }\href {\doibase
  10.1088/1674-1137/36/1/002, 10.1088/1674-1137/36/3/006} {\bibfield  {journal}
  {\bibinfo  {journal} {Chin. Phys.}\ }\textbf {\bibinfo {volume} {C36}},\
  \bibinfo {pages} {6} (\bibinfo {year} {2012})},\ \Eprint
  {http://arxiv.org/abs/1105.2901} {arXiv:1105.2901 [hep-ph]} \BibitemShut
  {NoStop}%
\bibitem [{\citenamefont {Yuan}\ \emph {et~al.}(2012)\citenamefont {Yuan},
  \citenamefont {Wei}, \citenamefont {He}, \citenamefont {Xu},\ and\
  \citenamefont {Zou}}]{Yuan:2012wz}%
  \BibitemOpen
  \bibfield  {author} {\bibinfo {author} {\bibfnamefont {S.~G.}\ \bibnamefont
  {Yuan}}, \bibinfo {author} {\bibfnamefont {K.~W.}\ \bibnamefont {Wei}},
  \bibinfo {author} {\bibfnamefont {J.}~\bibnamefont {He}}, \bibinfo {author}
  {\bibfnamefont {H.~S.}\ \bibnamefont {Xu}}, \ and\ \bibinfo {author}
  {\bibfnamefont {B.~S.}\ \bibnamefont {Zou}},\ }\href {\doibase
  10.1140/epja/i2012-12061-2} {\bibfield  {journal} {\bibinfo  {journal} {Eur.
  Phys. J.}\ }\textbf {\bibinfo {volume} {A48}},\ \bibinfo {pages} {61}
  (\bibinfo {year} {2012})},\ \Eprint {http://arxiv.org/abs/1201.0807}
  {arXiv:1201.0807 [nucl-th]} \BibitemShut {NoStop}%
\bibitem [{\citenamefont {Wu}\ \emph {et~al.}(2012)\citenamefont {Wu},
  \citenamefont {Lee},\ and\ \citenamefont {Zou}}]{Wu:2012md}%
  \BibitemOpen
  \bibfield  {author} {\bibinfo {author} {\bibfnamefont {J.-J.}\ \bibnamefont
  {Wu}}, \bibinfo {author} {\bibfnamefont {T.~S.~H.}\ \bibnamefont {Lee}}, \
  and\ \bibinfo {author} {\bibfnamefont {B.~S.}\ \bibnamefont {Zou}},\ }\href
  {\doibase 10.1103/PhysRevC.85.044002} {\bibfield  {journal} {\bibinfo
  {journal} {Phys. Rev.}\ }\textbf {\bibinfo {volume} {C85}},\ \bibinfo {pages}
  {044002} (\bibinfo {year} {2012})},\ \Eprint {http://arxiv.org/abs/1202.1036}
  {arXiv:1202.1036 [nucl-th]} \BibitemShut {NoStop}%
\bibitem [{\citenamefont {Xiao}\ \emph {et~al.}(2013)\citenamefont {Xiao},
  \citenamefont {Nieves},\ and\ \citenamefont {Oset}}]{Xiao:2013yca}%
  \BibitemOpen
  \bibfield  {author} {\bibinfo {author} {\bibfnamefont {C.~W.}\ \bibnamefont
  {Xiao}}, \bibinfo {author} {\bibfnamefont {J.}~\bibnamefont {Nieves}}, \ and\
  \bibinfo {author} {\bibfnamefont {E.}~\bibnamefont {Oset}},\ }\href {\doibase
  10.1103/PhysRevD.88.056012} {\bibfield  {journal} {\bibinfo  {journal} {Phys.
  Rev.}\ }\textbf {\bibinfo {volume} {D88}},\ \bibinfo {pages} {056012}
  (\bibinfo {year} {2013})},\ \Eprint {http://arxiv.org/abs/1304.5368}
  {arXiv:1304.5368 [hep-ph]} \BibitemShut {NoStop}%
\bibitem [{\citenamefont {Uchino}\ \emph {et~al.}(2016)\citenamefont {Uchino},
  \citenamefont {Liang},\ and\ \citenamefont {Oset}}]{Uchino:2015uha}%
  \BibitemOpen
  \bibfield  {author} {\bibinfo {author} {\bibfnamefont {T.}~\bibnamefont
  {Uchino}}, \bibinfo {author} {\bibfnamefont {W.-H.}\ \bibnamefont {Liang}}, \
  and\ \bibinfo {author} {\bibfnamefont {E.}~\bibnamefont {Oset}},\ }\href
  {\doibase 10.1140/epja/i2016-16043-0} {\bibfield  {journal} {\bibinfo
  {journal} {Eur. Phys. J.}\ }\textbf {\bibinfo {volume} {A52}},\ \bibinfo
  {pages} {43} (\bibinfo {year} {2016})},\ \Eprint
  {http://arxiv.org/abs/1504.05726} {arXiv:1504.05726 [hep-ph]} \BibitemShut
  {NoStop}%
\bibitem [{\citenamefont {Karliner}\ and\ \citenamefont
  {Rosner}(2015)}]{Karliner:2015ina}%
  \BibitemOpen
  \bibfield  {author} {\bibinfo {author} {\bibfnamefont {M.}~\bibnamefont
  {Karliner}}\ and\ \bibinfo {author} {\bibfnamefont {J.~L.}\ \bibnamefont
  {Rosner}},\ }\href {\doibase 10.1103/PhysRevLett.115.122001} {\bibfield
  {journal} {\bibinfo  {journal} {Phys. Rev. Lett.}\ }\textbf {\bibinfo
  {volume} {115}},\ \bibinfo {pages} {122001} (\bibinfo {year} {2015})},\
  \Eprint {http://arxiv.org/abs/1506.06386} {arXiv:1506.06386 [hep-ph]}
  \BibitemShut {NoStop}%
\bibitem [{\citenamefont {Xiao}\ \emph
  {et~al.}(2019{\natexlab{a}})\citenamefont {Xiao}, \citenamefont {Huang},
  \citenamefont {Dong}, \citenamefont {Geng},\ and\ \citenamefont
  {Chen}}]{Xiao:2019mst}%
  \BibitemOpen
  \bibfield  {author} {\bibinfo {author} {\bibfnamefont {C.-J.}\ \bibnamefont
  {Xiao}}, \bibinfo {author} {\bibfnamefont {Y.}~\bibnamefont {Huang}},
  \bibinfo {author} {\bibfnamefont {Y.-B.}\ \bibnamefont {Dong}}, \bibinfo
  {author} {\bibfnamefont {L.-S.}\ \bibnamefont {Geng}}, \ and\ \bibinfo
  {author} {\bibfnamefont {D.-Y.}\ \bibnamefont {Chen}},\ }\href {\doibase
  10.1103/PhysRevD.100.014022} {\bibfield  {journal} {\bibinfo  {journal}
  {Phys. Rev.}\ }\textbf {\bibinfo {volume} {D100}},\ \bibinfo {pages} {014022}
  (\bibinfo {year} {2019}{\natexlab{a}})},\ \Eprint
  {http://arxiv.org/abs/1904.00872} {arXiv:1904.00872 [hep-ph]} \BibitemShut
  {NoStop}%
\bibitem [{\citenamefont {Xiao}\ \emph
  {et~al.}(2019{\natexlab{b}})\citenamefont {Xiao}, \citenamefont {Nieves},\
  and\ \citenamefont {Oset}}]{Xiao:2019aya}%
  \BibitemOpen
  \bibfield  {author} {\bibinfo {author} {\bibfnamefont {C.~W.}\ \bibnamefont
  {Xiao}}, \bibinfo {author} {\bibfnamefont {J.}~\bibnamefont {Nieves}}, \ and\
  \bibinfo {author} {\bibfnamefont {E.}~\bibnamefont {Oset}},\ }\href {\doibase
  10.1103/PhysRevD.100.014021} {\bibfield  {journal} {\bibinfo  {journal}
  {Phys. Rev.}\ }\textbf {\bibinfo {volume} {D100}},\ \bibinfo {pages} {014021}
  (\bibinfo {year} {2019}{\natexlab{b}})},\ \Eprint
  {http://arxiv.org/abs/1904.01296} {arXiv:1904.01296 [hep-ph]} \BibitemShut
  {NoStop}%
\bibitem [{\citenamefont {Liu}\ \emph {et~al.}(2019{\natexlab{a}})\citenamefont
  {Liu}, \citenamefont {Pan}, \citenamefont {Peng}, \citenamefont
  {Sánchez~Sánchez}, \citenamefont {Geng}, \citenamefont {Hosaka},\ and\
  \citenamefont {Pavon~Valderrama}}]{Liu:2019tjn}%
  \BibitemOpen
  \bibfield  {author} {\bibinfo {author} {\bibfnamefont {M.-Z.}\ \bibnamefont
  {Liu}}, \bibinfo {author} {\bibfnamefont {Y.-W.}\ \bibnamefont {Pan}},
  \bibinfo {author} {\bibfnamefont {F.-Z.}\ \bibnamefont {Peng}}, \bibinfo
  {author} {\bibfnamefont {M.}~\bibnamefont {Sánchez~Sánchez}}, \bibinfo
  {author} {\bibfnamefont {L.-S.}\ \bibnamefont {Geng}}, \bibinfo {author}
  {\bibfnamefont {A.}~\bibnamefont {Hosaka}}, \ and\ \bibinfo {author}
  {\bibfnamefont {M.}~\bibnamefont {Pavon~Valderrama}},\ }\href {\doibase
  10.1103/PhysRevLett.122.242001} {\bibfield  {journal} {\bibinfo  {journal}
  {Phys. Rev. Lett.}\ }\textbf {\bibinfo {volume} {122}},\ \bibinfo {pages}
  {242001} (\bibinfo {year} {2019}{\natexlab{a}})},\ \Eprint
  {http://arxiv.org/abs/1903.11560} {arXiv:1903.11560 [hep-ph]} \BibitemShut
  {NoStop}%
\bibitem [{\citenamefont {Chen}\ \emph
  {et~al.}(2019{\natexlab{a}})\citenamefont {Chen}, \citenamefont {Chen},\ and\
  \citenamefont {Zhu}}]{Chen:2019bip}%
  \BibitemOpen
  \bibfield  {author} {\bibinfo {author} {\bibfnamefont {H.-X.}\ \bibnamefont
  {Chen}}, \bibinfo {author} {\bibfnamefont {W.}~\bibnamefont {Chen}}, \ and\
  \bibinfo {author} {\bibfnamefont {S.-L.}\ \bibnamefont {Zhu}},\ }\href
  {\doibase 10.1103/PhysRevD.100.051501} {\bibfield  {journal} {\bibinfo
  {journal} {Phys. Rev.}\ }\textbf {\bibinfo {volume} {D100}},\ \bibinfo
  {pages} {051501} (\bibinfo {year} {2019}{\natexlab{a}})},\ \Eprint
  {http://arxiv.org/abs/1903.11001} {arXiv:1903.11001 [hep-ph]} \BibitemShut
  {NoStop}%
\bibitem [{\citenamefont {Chen}\ \emph
  {et~al.}(2019{\natexlab{b}})\citenamefont {Chen}, \citenamefont {Sun},
  \citenamefont {Liu},\ and\ \citenamefont {Zhu}}]{Chen:2019asm}%
  \BibitemOpen
  \bibfield  {author} {\bibinfo {author} {\bibfnamefont {R.}~\bibnamefont
  {Chen}}, \bibinfo {author} {\bibfnamefont {Z.-F.}\ \bibnamefont {Sun}},
  \bibinfo {author} {\bibfnamefont {X.}~\bibnamefont {Liu}}, \ and\ \bibinfo
  {author} {\bibfnamefont {S.-L.}\ \bibnamefont {Zhu}},\ }\href {\doibase
  10.1103/PhysRevD.100.011502} {\bibfield  {journal} {\bibinfo  {journal}
  {Phys. Rev.}\ }\textbf {\bibinfo {volume} {D100}},\ \bibinfo {pages} {011502}
  (\bibinfo {year} {2019}{\natexlab{b}})},\ \Eprint
  {http://arxiv.org/abs/1903.11013} {arXiv:1903.11013 [hep-ph]} \BibitemShut
  {NoStop}%
\bibitem [{\citenamefont {Fernández-Ramírez}\ \emph
  {et~al.}(2019)\citenamefont {Fernández-Ramírez}, \citenamefont {Pilloni},
  \citenamefont {Albaladejo}, \citenamefont {Jackura}, \citenamefont {Mathieu},
  \citenamefont {Mikhasenko}, \citenamefont {Silva-Castro},\ and\ \citenamefont
  {Szczepaniak}}]{Fernandez-Ramirez:2019koa}%
  \BibitemOpen
  \bibfield  {author} {\bibinfo {author} {\bibfnamefont {C.}~\bibnamefont
  {Fernández-Ramírez}}, \bibinfo {author} {\bibfnamefont {A.}~\bibnamefont
  {Pilloni}}, \bibinfo {author} {\bibfnamefont {M.}~\bibnamefont {Albaladejo}},
  \bibinfo {author} {\bibfnamefont {A.}~\bibnamefont {Jackura}}, \bibinfo
  {author} {\bibfnamefont {V.}~\bibnamefont {Mathieu}}, \bibinfo {author}
  {\bibfnamefont {M.}~\bibnamefont {Mikhasenko}}, \bibinfo {author}
  {\bibfnamefont {J.~A.}\ \bibnamefont {Silva-Castro}}, \ and\ \bibinfo
  {author} {\bibfnamefont {A.~P.}\ \bibnamefont {Szczepaniak}} (\bibinfo
  {collaboration} {JPAC}),\ }\href {\doibase 10.1103/PhysRevLett.123.092001}
  {\bibfield  {journal} {\bibinfo  {journal} {Phys. Rev. Lett.}\ }\textbf
  {\bibinfo {volume} {123}},\ \bibinfo {pages} {092001} (\bibinfo {year}
  {2019})},\ \Eprint {http://arxiv.org/abs/1904.10021} {arXiv:1904.10021
  [hep-ph]} \BibitemShut {NoStop}%
\bibitem [{\citenamefont {Guo}\ \emph {et~al.}(2019)\citenamefont {Guo},
  \citenamefont {Jing}, \citenamefont {Meißner},\ and\ \citenamefont
  {Sakai}}]{Guo:2019fdo}%
  \BibitemOpen
  \bibfield  {author} {\bibinfo {author} {\bibfnamefont {F.-K.}\ \bibnamefont
  {Guo}}, \bibinfo {author} {\bibfnamefont {H.-J.}\ \bibnamefont {Jing}},
  \bibinfo {author} {\bibfnamefont {U.-G.}\ \bibnamefont {Meißner}}, \ and\
  \bibinfo {author} {\bibfnamefont {S.}~\bibnamefont {Sakai}},\ }\href
  {\doibase 10.1103/PhysRevD.99.091501} {\bibfield  {journal} {\bibinfo
  {journal} {Phys. Rev.}\ }\textbf {\bibinfo {volume} {D99}},\ \bibinfo {pages}
  {091501} (\bibinfo {year} {2019})},\ \Eprint
  {http://arxiv.org/abs/1903.11503} {arXiv:1903.11503 [hep-ph]} \BibitemShut
  {NoStop}%
\bibitem [{\citenamefont {He}(2019)}]{He:2019ify}%
  \BibitemOpen
  \bibfield  {author} {\bibinfo {author} {\bibfnamefont {J.}~\bibnamefont
  {He}},\ }\href {\doibase 10.1140/epjc/s10052-019-6906-1} {\bibfield
  {journal} {\bibinfo  {journal} {Eur. Phys. J.}\ }\textbf {\bibinfo {volume}
  {C79}},\ \bibinfo {pages} {393} (\bibinfo {year} {2019})},\ \Eprint
  {http://arxiv.org/abs/1903.11872} {arXiv:1903.11872 [hep-ph]} \BibitemShut
  {NoStop}%
\bibitem [{\citenamefont {Zhu}\ \emph {et~al.}(2019)\citenamefont {Zhu},
  \citenamefont {Liu}, \citenamefont {Huang},\ and\ \citenamefont
  {Qiao}}]{Zhu:2019iwm}%
  \BibitemOpen
  \bibfield  {author} {\bibinfo {author} {\bibfnamefont {R.}~\bibnamefont
  {Zhu}}, \bibinfo {author} {\bibfnamefont {X.}~\bibnamefont {Liu}}, \bibinfo
  {author} {\bibfnamefont {H.}~\bibnamefont {Huang}}, \ and\ \bibinfo {author}
  {\bibfnamefont {C.-F.}\ \bibnamefont {Qiao}},\ }\href {\doibase
  10.1016/j.physletb.2019.134869} {\bibfield  {journal} {\bibinfo  {journal}
  {Phys. Lett.}\ }\textbf {\bibinfo {volume} {B797}},\ \bibinfo {pages}
  {134869} (\bibinfo {year} {2019})},\ \Eprint
  {http://arxiv.org/abs/1904.10285} {arXiv:1904.10285 [hep-ph]} \BibitemShut
  {NoStop}%
\bibitem [{\citenamefont {Huang}\ \emph {et~al.}(2019)\citenamefont {Huang},
  \citenamefont {He},\ and\ \citenamefont {Ping}}]{Huang:2019jlf}%
  \BibitemOpen
  \bibfield  {author} {\bibinfo {author} {\bibfnamefont {H.}~\bibnamefont
  {Huang}}, \bibinfo {author} {\bibfnamefont {J.}~\bibnamefont {He}}, \ and\
  \bibinfo {author} {\bibfnamefont {J.}~\bibnamefont {Ping}},\ }\href@noop {}
  {\  (\bibinfo {year} {2019})},\ \Eprint {http://arxiv.org/abs/1904.00221}
  {arXiv:1904.00221 [hep-ph]} \BibitemShut {NoStop}%
\bibitem [{\citenamefont {Ali}\ and\ \citenamefont
  {Parkhomenko}(2019)}]{Ali:2019npk}%
  \BibitemOpen
  \bibfield  {author} {\bibinfo {author} {\bibfnamefont {A.}~\bibnamefont
  {Ali}}\ and\ \bibinfo {author} {\bibfnamefont {A.~{\relax Ya}.}\ \bibnamefont
  {Parkhomenko}},\ }\href {\doibase 10.1016/j.physletb.2019.05.002} {\bibfield
  {journal} {\bibinfo  {journal} {Phys. Lett.}\ }\textbf {\bibinfo {volume}
  {B793}},\ \bibinfo {pages} {365} (\bibinfo {year} {2019})},\ \Eprint
  {http://arxiv.org/abs/1904.00446} {arXiv:1904.00446 [hep-ph]} \BibitemShut
  {NoStop}%
\bibitem [{\citenamefont {Shimizu}\ \emph {et~al.}(2019)\citenamefont
  {Shimizu}, \citenamefont {Yamaguchi},\ and\ \citenamefont
  {Harada}}]{Shimizu:2019ptd}%
  \BibitemOpen
  \bibfield  {author} {\bibinfo {author} {\bibfnamefont {Y.}~\bibnamefont
  {Shimizu}}, \bibinfo {author} {\bibfnamefont {Y.}~\bibnamefont {Yamaguchi}},
  \ and\ \bibinfo {author} {\bibfnamefont {M.}~\bibnamefont {Harada}},\
  }\href@noop {} {\  (\bibinfo {year} {2019})},\ \Eprint
  {http://arxiv.org/abs/1904.00587} {arXiv:1904.00587 [hep-ph]} \BibitemShut
  {NoStop}%
\bibitem [{\citenamefont {Guo}\ and\ \citenamefont
  {Oller}(2019)}]{Guo:2019kdc}%
  \BibitemOpen
  \bibfield  {author} {\bibinfo {author} {\bibfnamefont {Z.-H.}\ \bibnamefont
  {Guo}}\ and\ \bibinfo {author} {\bibfnamefont {J.~A.}\ \bibnamefont
  {Oller}},\ }\href {\doibase 10.1016/j.physletb.2019.04.053} {\bibfield
  {journal} {\bibinfo  {journal} {Phys. Lett.}\ }\textbf {\bibinfo {volume}
  {B793}},\ \bibinfo {pages} {144} (\bibinfo {year} {2019})},\ \Eprint
  {http://arxiv.org/abs/1904.00851} {arXiv:1904.00851 [hep-ph]} \BibitemShut
  {NoStop}%
\bibitem [{\citenamefont {Mutuk}(2019)}]{Mutuk:2019snd}%
  \BibitemOpen
  \bibfield  {author} {\bibinfo {author} {\bibfnamefont {H.}~\bibnamefont
  {Mutuk}},\ }\href {\doibase 10.1088/1674-1137/43/9/093103} {\bibfield
  {journal} {\bibinfo  {journal} {Chin. Phys.}\ }\textbf {\bibinfo {volume}
  {C43}},\ \bibinfo {pages} {093103} (\bibinfo {year} {2019})},\ \Eprint
  {http://arxiv.org/abs/1904.09756} {arXiv:1904.09756 [hep-ph]} \BibitemShut
  {NoStop}%
\bibitem [{\citenamefont {Weng}\ \emph {et~al.}(2019)\citenamefont {Weng},
  \citenamefont {Chen}, \citenamefont {Deng},\ and\ \citenamefont
  {Zhu}}]{Weng:2019ynv}%
  \BibitemOpen
  \bibfield  {author} {\bibinfo {author} {\bibfnamefont {X.-Z.}\ \bibnamefont
  {Weng}}, \bibinfo {author} {\bibfnamefont {X.-L.}\ \bibnamefont {Chen}},
  \bibinfo {author} {\bibfnamefont {W.-Z.}\ \bibnamefont {Deng}}, \ and\
  \bibinfo {author} {\bibfnamefont {S.-L.}\ \bibnamefont {Zhu}},\ }\href
  {\doibase 10.1103/PhysRevD.100.016014} {\bibfield  {journal} {\bibinfo
  {journal} {Phys. Rev.}\ }\textbf {\bibinfo {volume} {D100}},\ \bibinfo
  {pages} {016014} (\bibinfo {year} {2019})},\ \Eprint
  {http://arxiv.org/abs/1904.09891} {arXiv:1904.09891 [hep-ph]} \BibitemShut
  {NoStop}%
\bibitem [{\citenamefont {Eides}\ \emph {et~al.}(2019)\citenamefont {Eides},
  \citenamefont {Petrov},\ and\ \citenamefont {Polyakov}}]{Eides:2019tgv}%
  \BibitemOpen
  \bibfield  {author} {\bibinfo {author} {\bibfnamefont {M.~I.}\ \bibnamefont
  {Eides}}, \bibinfo {author} {\bibfnamefont {V.~Y.}\ \bibnamefont {Petrov}}, \
  and\ \bibinfo {author} {\bibfnamefont {M.~V.}\ \bibnamefont {Polyakov}},\
  }\href@noop {} {\  (\bibinfo {year} {2019})},\ \Eprint
  {http://arxiv.org/abs/1904.11616} {arXiv:1904.11616 [hep-ph]} \BibitemShut
  {NoStop}%
\bibitem [{\citenamefont {Wang}(2019{\natexlab{a}})}]{Wang:2018waa}%
  \BibitemOpen
  \bibfield  {author} {\bibinfo {author} {\bibfnamefont {Z.-G.}\ \bibnamefont
  {Wang}},\ }\href {\doibase 10.1142/S0217751X19500970} {\bibfield  {journal}
  {\bibinfo  {journal} {Int. J. Mod. Phys.}\ }\textbf {\bibinfo {volume}
  {A34}},\ \bibinfo {pages} {1950097} (\bibinfo {year} {2019}{\natexlab{a}})},\
  \Eprint {http://arxiv.org/abs/1806.10384} {arXiv:1806.10384 [hep-ph]}
  \BibitemShut {NoStop}%
\bibitem [{\citenamefont {Wang}(2019{\natexlab{b}})}]{Wang:2019got}%
  \BibitemOpen
  \bibfield  {author} {\bibinfo {author} {\bibfnamefont {Z.-G.}\ \bibnamefont
  {Wang}},\ }\href@noop {} {\  (\bibinfo {year} {2019}{\natexlab{b}})},\
  \Eprint {http://arxiv.org/abs/1905.02892} {arXiv:1905.02892 [hep-ph]}
  \BibitemShut {NoStop}%
\bibitem [{\citenamefont {Meng}\ \emph {et~al.}(2019)\citenamefont {Meng},
  \citenamefont {Wang}, \citenamefont {Wang},\ and\ \citenamefont
  {Zhu}}]{Meng:2019ilv}%
  \BibitemOpen
  \bibfield  {author} {\bibinfo {author} {\bibfnamefont {L.}~\bibnamefont
  {Meng}}, \bibinfo {author} {\bibfnamefont {B.}~\bibnamefont {Wang}}, \bibinfo
  {author} {\bibfnamefont {G.-J.}\ \bibnamefont {Wang}}, \ and\ \bibinfo
  {author} {\bibfnamefont {S.-L.}\ \bibnamefont {Zhu}},\ }\href {\doibase
  10.1103/PhysRevD.100.014031} {\bibfield  {journal} {\bibinfo  {journal}
  {Phys. Rev.}\ }\textbf {\bibinfo {volume} {D100}},\ \bibinfo {pages} {014031}
  (\bibinfo {year} {2019})},\ \Eprint {http://arxiv.org/abs/1905.04113}
  {arXiv:1905.04113 [hep-ph]} \BibitemShut {NoStop}%
\bibitem [{\citenamefont {Cheng}\ and\ \citenamefont
  {Liu}(2019)}]{Cheng:2019obk}%
  \BibitemOpen
  \bibfield  {author} {\bibinfo {author} {\bibfnamefont {J.-B.}\ \bibnamefont
  {Cheng}}\ and\ \bibinfo {author} {\bibfnamefont {Y.-R.}\ \bibnamefont
  {Liu}},\ }\href {\doibase 10.1103/PhysRevD.100.054002} {\bibfield  {journal}
  {\bibinfo  {journal} {Phys. Rev.}\ }\textbf {\bibinfo {volume} {D100}},\
  \bibinfo {pages} {054002} (\bibinfo {year} {2019})},\ \Eprint
  {http://arxiv.org/abs/1905.08605} {arXiv:1905.08605 [hep-ph]} \BibitemShut
  {NoStop}%
\bibitem [{\citenamefont {Wang}\ \emph
  {et~al.}(2019{\natexlab{a}})\citenamefont {Wang}, \citenamefont {Chen},
  \citenamefont {Liu},\ and\ \citenamefont {Liu}}]{Wang:2019nwt}%
  \BibitemOpen
  \bibfield  {author} {\bibinfo {author} {\bibfnamefont {F.-L.}\ \bibnamefont
  {Wang}}, \bibinfo {author} {\bibfnamefont {R.}~\bibnamefont {Chen}}, \bibinfo
  {author} {\bibfnamefont {Z.-W.}\ \bibnamefont {Liu}}, \ and\ \bibinfo
  {author} {\bibfnamefont {X.}~\bibnamefont {Liu}},\ }\href@noop {} {\
  (\bibinfo {year} {2019}{\natexlab{a}})},\ \Eprint
  {http://arxiv.org/abs/1905.03636} {arXiv:1905.03636 [hep-ph]} \BibitemShut
  {NoStop}%
\bibitem [{\citenamefont {Wu}\ and\ \citenamefont {Chen}(2019)}]{Wu:2019rog}%
  \BibitemOpen
  \bibfield  {author} {\bibinfo {author} {\bibfnamefont {Q.}~\bibnamefont
  {Wu}}\ and\ \bibinfo {author} {\bibfnamefont {D.-Y.}\ \bibnamefont {Chen}},\
  }\href@noop {} {\  (\bibinfo {year} {2019})},\ \Eprint
  {http://arxiv.org/abs/1906.02480} {arXiv:1906.02480 [hep-ph]} \BibitemShut
  {NoStop}%
\bibitem [{\citenamefont {Wang}\ \emph
  {et~al.}(2019{\natexlab{b}})\citenamefont {Wang}, \citenamefont {Chen},\ and\
  \citenamefont {He}}]{Wang:2019krd}%
  \BibitemOpen
  \bibfield  {author} {\bibinfo {author} {\bibfnamefont {X.-Y.}\ \bibnamefont
  {Wang}}, \bibinfo {author} {\bibfnamefont {X.-R.}\ \bibnamefont {Chen}}, \
  and\ \bibinfo {author} {\bibfnamefont {J.}~\bibnamefont {He}},\ }\href
  {\doibase 10.1103/PhysRevD.99.114007} {\bibfield  {journal} {\bibinfo
  {journal} {Phys. Rev.}\ }\textbf {\bibinfo {volume} {D99}},\ \bibinfo {pages}
  {114007} (\bibinfo {year} {2019}{\natexlab{b}})},\ \Eprint
  {http://arxiv.org/abs/1904.11706} {arXiv:1904.11706 [hep-ph]} \BibitemShut
  {NoStop}%
\bibitem [{\citenamefont {Cao}\ and\ \citenamefont {Dai}(2019)}]{Cao:2019kst}%
  \BibitemOpen
  \bibfield  {author} {\bibinfo {author} {\bibfnamefont {X.}~\bibnamefont
  {Cao}}\ and\ \bibinfo {author} {\bibfnamefont {J.-p.}\ \bibnamefont {Dai}},\
  }\href {\doibase 10.1103/PhysRevD.100.054033} {\bibfield  {journal} {\bibinfo
   {journal} {Phys. Rev.}\ }\textbf {\bibinfo {volume} {D100}},\ \bibinfo
  {pages} {054033} (\bibinfo {year} {2019})},\ \Eprint
  {http://arxiv.org/abs/1904.06015} {arXiv:1904.06015 [hep-ph]} \BibitemShut
  {NoStop}%
\bibitem [{\citenamefont {Ali}\ \emph {et~al.}(2019)\citenamefont {Ali} \emph
  {et~al.}}]{Ali:2019lzf}%
  \BibitemOpen
  \bibfield  {author} {\bibinfo {author} {\bibfnamefont {A.}~\bibnamefont
  {Ali}} \emph {et~al.} (\bibinfo {collaboration} {GlueX}),\ }\href {\doibase
  10.1103/PhysRevLett.123.072001} {\bibfield  {journal} {\bibinfo  {journal}
  {Phys. Rev. Lett.}\ }\textbf {\bibinfo {volume} {123}},\ \bibinfo {pages}
  {072001} (\bibinfo {year} {2019})},\ \Eprint
  {http://arxiv.org/abs/1905.10811} {arXiv:1905.10811 [nucl-ex]} \BibitemShut
  {NoStop}%
\bibitem [{\citenamefont {Wang}\ \emph
  {et~al.}(2019{\natexlab{c}})\citenamefont {Wang}, \citenamefont {He},
  \citenamefont {Chen}, \citenamefont {Wang},\ and\ \citenamefont
  {Zhu}}]{Wang:2019dsi}%
  \BibitemOpen
  \bibfield  {author} {\bibinfo {author} {\bibfnamefont {X.-Y.}\ \bibnamefont
  {Wang}}, \bibinfo {author} {\bibfnamefont {J.}~\bibnamefont {He}}, \bibinfo
  {author} {\bibfnamefont {X.-R.}\ \bibnamefont {Chen}}, \bibinfo {author}
  {\bibfnamefont {Q.}~\bibnamefont {Wang}}, \ and\ \bibinfo {author}
  {\bibfnamefont {X.}~\bibnamefont {Zhu}},\ }\href {\doibase
  10.1016/j.physletb.2019.134862} {\bibfield  {journal} {\bibinfo  {journal}
  {Phys. Lett.}\ }\textbf {\bibinfo {volume} {B797}},\ \bibinfo {pages}
  {134862} (\bibinfo {year} {2019}{\natexlab{c}})},\ \Eprint
  {http://arxiv.org/abs/1906.04044} {arXiv:1906.04044 [hep-ph]} \BibitemShut
  {NoStop}%
\bibitem [{\citenamefont {Wu}\ \emph {et~al.}(2019)\citenamefont {Wu},
  \citenamefont {Lee},\ and\ \citenamefont {Zou}}]{Wu:2019adv}%
  \BibitemOpen
  \bibfield  {author} {\bibinfo {author} {\bibfnamefont {J.-J.}\ \bibnamefont
  {Wu}}, \bibinfo {author} {\bibfnamefont {T.~S.~H.}\ \bibnamefont {Lee}}, \
  and\ \bibinfo {author} {\bibfnamefont {B.-S.}\ \bibnamefont {Zou}},\ }\href
  {\doibase 10.1103/PhysRevC.100.035206} {\bibfield  {journal} {\bibinfo
  {journal} {Phys. Rev.}\ }\textbf {\bibinfo {volume} {C100}},\ \bibinfo
  {pages} {035206} (\bibinfo {year} {2019})},\ \Eprint
  {http://arxiv.org/abs/1906.05375} {arXiv:1906.05375 [nucl-th]} \BibitemShut
  {NoStop}%
\bibitem [{\citenamefont {Holma}\ and\ \citenamefont
  {Ohlsson}(2019)}]{Holma:2019lxe}%
  \BibitemOpen
  \bibfield  {author} {\bibinfo {author} {\bibfnamefont {P.}~\bibnamefont
  {Holma}}\ and\ \bibinfo {author} {\bibfnamefont {T.}~\bibnamefont
  {Ohlsson}},\ }\href@noop {} {\  (\bibinfo {year} {2019})},\ \Eprint
  {http://arxiv.org/abs/1906.08499} {arXiv:1906.08499 [hep-ph]} \BibitemShut
  {NoStop}%
\bibitem [{\citenamefont {Yamaguchi}\ \emph {et~al.}(2019)\citenamefont
  {Yamaguchi}, \citenamefont {García-Tecocoatzi}, \citenamefont {Giachino},
  \citenamefont {Hosaka}, \citenamefont {Santopinto}, \citenamefont
  {Takeuchi},\ and\ \citenamefont {Takizawa}}]{Yamaguchi:2019seo}%
  \BibitemOpen
  \bibfield  {author} {\bibinfo {author} {\bibfnamefont {Y.}~\bibnamefont
  {Yamaguchi}}, \bibinfo {author} {\bibfnamefont {H.}~\bibnamefont
  {García-Tecocoatzi}}, \bibinfo {author} {\bibfnamefont {A.}~\bibnamefont
  {Giachino}}, \bibinfo {author} {\bibfnamefont {A.}~\bibnamefont {Hosaka}},
  \bibinfo {author} {\bibfnamefont {E.}~\bibnamefont {Santopinto}}, \bibinfo
  {author} {\bibfnamefont {S.}~\bibnamefont {Takeuchi}}, \ and\ \bibinfo
  {author} {\bibfnamefont {M.}~\bibnamefont {Takizawa}},\ }\href@noop {} {\
  (\bibinfo {year} {2019})},\ \Eprint {http://arxiv.org/abs/1907.04684}
  {arXiv:1907.04684 [hep-ph]} \BibitemShut {NoStop}%
\bibitem [{\citenamefont {Chen}\ \emph {et~al.}(2016)\citenamefont {Chen},
  \citenamefont {Chen}, \citenamefont {Liu},\ and\ \citenamefont
  {Zhu}}]{Chen:2016qju}%
  \BibitemOpen
  \bibfield  {author} {\bibinfo {author} {\bibfnamefont {H.-X.}\ \bibnamefont
  {Chen}}, \bibinfo {author} {\bibfnamefont {W.}~\bibnamefont {Chen}}, \bibinfo
  {author} {\bibfnamefont {X.}~\bibnamefont {Liu}}, \ and\ \bibinfo {author}
  {\bibfnamefont {S.-L.}\ \bibnamefont {Zhu}},\ }\href {\doibase
  10.1016/j.physrep.2016.05.004} {\bibfield  {journal} {\bibinfo  {journal}
  {Phys. Rept.}\ }\textbf {\bibinfo {volume} {639}},\ \bibinfo {pages} {1}
  (\bibinfo {year} {2016})},\ \Eprint {http://arxiv.org/abs/1601.02092}
  {arXiv:1601.02092 [hep-ph]} \BibitemShut {NoStop}%
\bibitem [{\citenamefont {Lebed}\ \emph {et~al.}(2017)\citenamefont {Lebed},
  \citenamefont {Mitchell},\ and\ \citenamefont {Swanson}}]{Lebed:2016hpi}%
  \BibitemOpen
  \bibfield  {author} {\bibinfo {author} {\bibfnamefont {R.~F.}\ \bibnamefont
  {Lebed}}, \bibinfo {author} {\bibfnamefont {R.~E.}\ \bibnamefont {Mitchell}},
  \ and\ \bibinfo {author} {\bibfnamefont {E.~S.}\ \bibnamefont {Swanson}},\
  }\href {\doibase 10.1016/j.ppnp.2016.11.003} {\bibfield  {journal} {\bibinfo
  {journal} {Prog. Part. Nucl. Phys.}\ }\textbf {\bibinfo {volume} {93}},\
  \bibinfo {pages} {143} (\bibinfo {year} {2017})},\ \Eprint
  {http://arxiv.org/abs/1610.04528} {arXiv:1610.04528 [hep-ph]} \BibitemShut
  {NoStop}%
\bibitem [{\citenamefont {Esposito}\ \emph {et~al.}(2017)\citenamefont
  {Esposito}, \citenamefont {Pilloni},\ and\ \citenamefont
  {Polosa}}]{Esposito:2016noz}%
  \BibitemOpen
  \bibfield  {author} {\bibinfo {author} {\bibfnamefont {A.}~\bibnamefont
  {Esposito}}, \bibinfo {author} {\bibfnamefont {A.}~\bibnamefont {Pilloni}}, \
  and\ \bibinfo {author} {\bibfnamefont {A.~D.}\ \bibnamefont {Polosa}},\
  }\href {\doibase 10.1016/j.physrep.2016.11.002} {\bibfield  {journal}
  {\bibinfo  {journal} {Phys. Rept.}\ }\textbf {\bibinfo {volume} {668}},\
  \bibinfo {pages} {1} (\bibinfo {year} {2017})},\ \Eprint
  {http://arxiv.org/abs/1611.07920} {arXiv:1611.07920 [hep-ph]} \BibitemShut
  {NoStop}%
\bibitem [{\citenamefont {Ali}\ \emph {et~al.}(2017)\citenamefont {Ali},
  \citenamefont {Lange},\ and\ \citenamefont {Stone}}]{Ali:2017jda}%
  \BibitemOpen
  \bibfield  {author} {\bibinfo {author} {\bibfnamefont {A.}~\bibnamefont
  {Ali}}, \bibinfo {author} {\bibfnamefont {J.~S.}\ \bibnamefont {Lange}}, \
  and\ \bibinfo {author} {\bibfnamefont {S.}~\bibnamefont {Stone}},\ }\href
  {\doibase 10.1016/j.ppnp.2017.08.003} {\bibfield  {journal} {\bibinfo
  {journal} {Prog. Part. Nucl. Phys.}\ }\textbf {\bibinfo {volume} {97}},\
  \bibinfo {pages} {123} (\bibinfo {year} {2017})},\ \Eprint
  {http://arxiv.org/abs/1706.00610} {arXiv:1706.00610 [hep-ph]} \BibitemShut
  {NoStop}%
\bibitem [{\citenamefont {Olsen}\ \emph {et~al.}(2018)\citenamefont {Olsen},
  \citenamefont {Skwarnicki},\ and\ \citenamefont {Zieminska}}]{Olsen:2017bmm}%
  \BibitemOpen
  \bibfield  {author} {\bibinfo {author} {\bibfnamefont {S.~L.}\ \bibnamefont
  {Olsen}}, \bibinfo {author} {\bibfnamefont {T.}~\bibnamefont {Skwarnicki}}, \
  and\ \bibinfo {author} {\bibfnamefont {D.}~\bibnamefont {Zieminska}},\ }\href
  {\doibase 10.1103/RevModPhys.90.015003} {\bibfield  {journal} {\bibinfo
  {journal} {Rev. Mod. Phys.}\ }\textbf {\bibinfo {volume} {90}},\ \bibinfo
  {pages} {015003} (\bibinfo {year} {2018})},\ \Eprint
  {http://arxiv.org/abs/1708.04012} {arXiv:1708.04012 [hep-ph]} \BibitemShut
  {NoStop}%
\bibitem [{\citenamefont {Karliner}\ \emph {et~al.}(2018)\citenamefont
  {Karliner}, \citenamefont {Rosner},\ and\ \citenamefont
  {Skwarnicki}}]{Karliner:2017qhf}%
  \BibitemOpen
  \bibfield  {author} {\bibinfo {author} {\bibfnamefont {M.}~\bibnamefont
  {Karliner}}, \bibinfo {author} {\bibfnamefont {J.~L.}\ \bibnamefont
  {Rosner}}, \ and\ \bibinfo {author} {\bibfnamefont {T.}~\bibnamefont
  {Skwarnicki}},\ }\href {\doibase 10.1146/annurev-nucl-101917-020902}
  {\bibfield  {journal} {\bibinfo  {journal} {Ann. Rev. Nucl. Part. Sci.}\
  }\textbf {\bibinfo {volume} {68}},\ \bibinfo {pages} {17} (\bibinfo {year}
  {2018})},\ \Eprint {http://arxiv.org/abs/1711.10626} {arXiv:1711.10626
  [hep-ph]} \BibitemShut {NoStop}%
\bibitem [{\citenamefont {Cerri}\ \emph {et~al.}(2018)\citenamefont {Cerri}
  \emph {et~al.}}]{Cerri:2018ypt}%
  \BibitemOpen
  \bibfield  {author} {\bibinfo {author} {\bibfnamefont {A.}~\bibnamefont
  {Cerri}} \emph {et~al.},\ }\href@noop {} {\  (\bibinfo {year} {2018})},\
  \Eprint {http://arxiv.org/abs/1812.07638} {arXiv:1812.07638 [hep-ph]}
  \BibitemShut {NoStop}%
\bibitem [{\citenamefont {Liu}\ \emph {et~al.}(2019{\natexlab{b}})\citenamefont
  {Liu}, \citenamefont {Chen}, \citenamefont {Chen}, \citenamefont {Liu},\ and\
  \citenamefont {Zhu}}]{Liu:2019zoy}%
  \BibitemOpen
  \bibfield  {author} {\bibinfo {author} {\bibfnamefont {Y.-R.}\ \bibnamefont
  {Liu}}, \bibinfo {author} {\bibfnamefont {H.-X.}\ \bibnamefont {Chen}},
  \bibinfo {author} {\bibfnamefont {W.}~\bibnamefont {Chen}}, \bibinfo {author}
  {\bibfnamefont {X.}~\bibnamefont {Liu}}, \ and\ \bibinfo {author}
  {\bibfnamefont {S.-L.}\ \bibnamefont {Zhu}},\ }\href {\doibase
  10.1016/j.ppnp.2019.04.003} {\bibfield  {journal} {\bibinfo  {journal} {Prog.
  Part. Nucl. Phys.}\ }\textbf {\bibinfo {volume} {107}},\ \bibinfo {pages}
  {237} (\bibinfo {year} {2019}{\natexlab{b}})},\ \Eprint
  {http://arxiv.org/abs/1903.11976} {arXiv:1903.11976 [hep-ph]} \BibitemShut
  {NoStop}%
\bibitem [{\citenamefont {Isgur}\ and\ \citenamefont
  {Wise}(1991)}]{Isgur:1991wq}%
  \BibitemOpen
  \bibfield  {author} {\bibinfo {author} {\bibfnamefont {N.}~\bibnamefont
  {Isgur}}\ and\ \bibinfo {author} {\bibfnamefont {M.~B.}\ \bibnamefont
  {Wise}},\ }\href {\doibase 10.1103/PhysRevLett.66.1130} {\bibfield  {journal}
  {\bibinfo  {journal} {Phys. Rev. Lett.}\ }\textbf {\bibinfo {volume} {66}},\
  \bibinfo {pages} {1130} (\bibinfo {year} {1991})}\BibitemShut {NoStop}%
\bibitem [{\citenamefont {Wise}(1992)}]{Wise:1992hn}%
  \BibitemOpen
  \bibfield  {author} {\bibinfo {author} {\bibfnamefont {M.~B.}\ \bibnamefont
  {Wise}},\ }\href {\doibase 10.1103/PhysRevD.45.R2188} {\bibfield  {journal}
  {\bibinfo  {journal} {Phys. Rev.}\ }\textbf {\bibinfo {volume} {D45}},\
  \bibinfo {pages} {R2188} (\bibinfo {year} {1992})}\BibitemShut {NoStop}%
\bibitem [{\citenamefont {Neubert}(1994)}]{Neubert:1993mb}%
  \BibitemOpen
  \bibfield  {author} {\bibinfo {author} {\bibfnamefont {M.}~\bibnamefont
  {Neubert}},\ }\href {\doibase 10.1016/0370-1573(94)90091-4} {\bibfield
  {journal} {\bibinfo  {journal} {Phys. Rept.}\ }\textbf {\bibinfo {volume}
  {245}},\ \bibinfo {pages} {259} (\bibinfo {year} {1994})},\ \Eprint
  {http://arxiv.org/abs/hep-ph/9306320} {arXiv:hep-ph/9306320 [hep-ph]}
  \BibitemShut {NoStop}%
\bibitem [{\citenamefont {Cleven}\ \emph {et~al.}(2015)\citenamefont {Cleven},
  \citenamefont {Guo}, \citenamefont {Hanhart}, \citenamefont {Wang},\ and\
  \citenamefont {Zhao}}]{Cleven:2015era}%
  \BibitemOpen
  \bibfield  {author} {\bibinfo {author} {\bibfnamefont {M.}~\bibnamefont
  {Cleven}}, \bibinfo {author} {\bibfnamefont {F.-K.}\ \bibnamefont {Guo}},
  \bibinfo {author} {\bibfnamefont {C.}~\bibnamefont {Hanhart}}, \bibinfo
  {author} {\bibfnamefont {Q.}~\bibnamefont {Wang}}, \ and\ \bibinfo {author}
  {\bibfnamefont {Q.}~\bibnamefont {Zhao}},\ }\href {\doibase
  10.1103/PhysRevD.92.014005} {\bibfield  {journal} {\bibinfo  {journal} {Phys.
  Rev.}\ }\textbf {\bibinfo {volume} {D92}},\ \bibinfo {pages} {014005}
  (\bibinfo {year} {2015})},\ \Eprint {http://arxiv.org/abs/1505.01771}
  {arXiv:1505.01771 [hep-ph]} \BibitemShut {NoStop}%
\bibitem [{\citenamefont {Xiao}\ \emph
  {et~al.}(2019{\natexlab{c}})\citenamefont {Xiao}, \citenamefont {Nieves},\
  and\ \citenamefont {Oset}}]{Xiao:2019gjd}%
  \BibitemOpen
  \bibfield  {author} {\bibinfo {author} {\bibfnamefont {C.~W.}\ \bibnamefont
  {Xiao}}, \bibinfo {author} {\bibfnamefont {J.}~\bibnamefont {Nieves}}, \ and\
  \bibinfo {author} {\bibfnamefont {E.}~\bibnamefont {Oset}},\ }\href@noop {}
  {\  (\bibinfo {year} {2019}{\natexlab{c}})},\ \Eprint
  {http://arxiv.org/abs/1906.09010} {arXiv:1906.09010 [hep-ph]} \BibitemShut
  {NoStop}%
\bibitem [{\citenamefont {Liu}\ \emph {et~al.}(2018)\citenamefont {Liu},
  \citenamefont {Peng}, \citenamefont {Sánchez~Sánchez},\ and\ \citenamefont
  {Valderrama}}]{Liu:2018zzu}%
  \BibitemOpen
  \bibfield  {author} {\bibinfo {author} {\bibfnamefont {M.-Z.}\ \bibnamefont
  {Liu}}, \bibinfo {author} {\bibfnamefont {F.-Z.}\ \bibnamefont {Peng}},
  \bibinfo {author} {\bibfnamefont {M.}~\bibnamefont {Sánchez~Sánchez}}, \
  and\ \bibinfo {author} {\bibfnamefont {M.~P.}\ \bibnamefont {Valderrama}},\
  }\href {\doibase 10.1103/PhysRevD.98.114030} {\bibfield  {journal} {\bibinfo
  {journal} {Phys. Rev.}\ }\textbf {\bibinfo {volume} {D98}},\ \bibinfo {pages}
  {114030} (\bibinfo {year} {2018})},\ \Eprint
  {http://arxiv.org/abs/1811.03992} {arXiv:1811.03992 [hep-ph]} \BibitemShut
  {NoStop}%
\bibitem [{\citenamefont {Shen}\ \emph {et~al.}(2016)\citenamefont {Shen},
  \citenamefont {Guo}, \citenamefont {Xie},\ and\ \citenamefont
  {Zou}}]{Shen:2016tzq}%
  \BibitemOpen
  \bibfield  {author} {\bibinfo {author} {\bibfnamefont {C.-W.}\ \bibnamefont
  {Shen}}, \bibinfo {author} {\bibfnamefont {F.-K.}\ \bibnamefont {Guo}},
  \bibinfo {author} {\bibfnamefont {J.-J.}\ \bibnamefont {Xie}}, \ and\
  \bibinfo {author} {\bibfnamefont {B.-S.}\ \bibnamefont {Zou}},\ }\href
  {\doibase 10.1016/j.nuclphysa.2016.04.034} {\bibfield  {journal} {\bibinfo
  {journal} {Nucl. Phys.}\ }\textbf {\bibinfo {volume} {A954}},\ \bibinfo
  {pages} {393} (\bibinfo {year} {2016})},\ \Eprint
  {http://arxiv.org/abs/1603.04672} {arXiv:1603.04672 [hep-ph]} \BibitemShut
  {NoStop}%
\bibitem [{\citenamefont {Lin}\ \emph {et~al.}(2017)\citenamefont {Lin},
  \citenamefont {Shen}, \citenamefont {Guo},\ and\ \citenamefont
  {Zou}}]{Lin:2017mtz}%
  \BibitemOpen
  \bibfield  {author} {\bibinfo {author} {\bibfnamefont {Y.-H.}\ \bibnamefont
  {Lin}}, \bibinfo {author} {\bibfnamefont {C.-W.}\ \bibnamefont {Shen}},
  \bibinfo {author} {\bibfnamefont {F.-K.}\ \bibnamefont {Guo}}, \ and\
  \bibinfo {author} {\bibfnamefont {B.-S.}\ \bibnamefont {Zou}},\ }\href
  {\doibase 10.1103/PhysRevD.95.114017} {\bibfield  {journal} {\bibinfo
  {journal} {Phys. Rev.}\ }\textbf {\bibinfo {volume} {D95}},\ \bibinfo {pages}
  {114017} (\bibinfo {year} {2017})},\ \Eprint
  {http://arxiv.org/abs/1703.01045} {arXiv:1703.01045 [hep-ph]} \BibitemShut
  {NoStop}%
\bibitem [{\citenamefont {Lin}\ and\ \citenamefont {Zou}(2019)}]{Lin:2019qiv}%
  \BibitemOpen
  \bibfield  {author} {\bibinfo {author} {\bibfnamefont {Y.-H.}\ \bibnamefont
  {Lin}}\ and\ \bibinfo {author} {\bibfnamefont {B.-S.}\ \bibnamefont {Zou}},\
  }\href {\doibase 10.1103/PhysRevD.100.056005} {\bibfield  {journal} {\bibinfo
   {journal} {Phys. Rev.}\ }\textbf {\bibinfo {volume} {D100}},\ \bibinfo
  {pages} {056005} (\bibinfo {year} {2019})},\ \Eprint
  {http://arxiv.org/abs/1908.05309} {arXiv:1908.05309 [hep-ph]} \BibitemShut
  {NoStop}%
\bibitem [{\citenamefont {Voloshin}(2019)}]{Voloshin:2019aut}%
  \BibitemOpen
  \bibfield  {author} {\bibinfo {author} {\bibfnamefont {M.~B.}\ \bibnamefont
  {Voloshin}},\ }\href {\doibase 10.1103/PhysRevD.100.034020} {\bibfield
  {journal} {\bibinfo  {journal} {Phys. Rev.}\ }\textbf {\bibinfo {volume}
  {D100}},\ \bibinfo {pages} {034020} (\bibinfo {year} {2019})},\ \Eprint
  {http://arxiv.org/abs/1907.01476} {arXiv:1907.01476 [hep-ph]} \BibitemShut
  {NoStop}%
\bibitem [{\citenamefont {Tanabashi}\ \emph {et~al.}(2018)\citenamefont
  {Tanabashi} \emph {et~al.}}]{Tanabashi:2018oca}%
  \BibitemOpen
  \bibfield  {author} {\bibinfo {author} {\bibfnamefont {M.}~\bibnamefont
  {Tanabashi}} \emph {et~al.} (\bibinfo {collaboration} {Particle Data
  Group}),\ }\href {\doibase 10.1103/PhysRevD.98.030001} {\bibfield  {journal}
  {\bibinfo  {journal} {Phys. Rev.}\ }\textbf {\bibinfo {volume} {D98}},\
  \bibinfo {pages} {030001} (\bibinfo {year} {2018})}\BibitemShut {NoStop}%
\bibitem [{\citenamefont {Guo}\ \emph {et~al.}(2006)\citenamefont {Guo},
  \citenamefont {Shen}, \citenamefont {Chiang}, \citenamefont {Ping},\ and\
  \citenamefont {Zou}}]{Guo:2006fu}%
  \BibitemOpen
  \bibfield  {author} {\bibinfo {author} {\bibfnamefont {F.-K.}\ \bibnamefont
  {Guo}}, \bibinfo {author} {\bibfnamefont {P.-N.}\ \bibnamefont {Shen}},
  \bibinfo {author} {\bibfnamefont {H.-C.}\ \bibnamefont {Chiang}}, \bibinfo
  {author} {\bibfnamefont {R.-G.}\ \bibnamefont {Ping}}, \ and\ \bibinfo
  {author} {\bibfnamefont {B.-S.}\ \bibnamefont {Zou}},\ }\href {\doibase
  10.1016/j.physletb.2006.08.064} {\bibfield  {journal} {\bibinfo  {journal}
  {Phys. Lett.}\ }\textbf {\bibinfo {volume} {B641}},\ \bibinfo {pages} {278}
  (\bibinfo {year} {2006})},\ \Eprint {http://arxiv.org/abs/hep-ph/0603072}
  {arXiv:hep-ph/0603072 [hep-ph]} \BibitemShut {NoStop}%
\bibitem [{\citenamefont {Casalbuoni}\ \emph {et~al.}(1997)\citenamefont
  {Casalbuoni}, \citenamefont {Deandrea}, \citenamefont {Di~Bartolomeo},
  \citenamefont {Gatto}, \citenamefont {Feruglio},\ and\ \citenamefont
  {Nardulli}}]{Casalbuoni:1996pg}%
  \BibitemOpen
  \bibfield  {author} {\bibinfo {author} {\bibfnamefont {R.}~\bibnamefont
  {Casalbuoni}}, \bibinfo {author} {\bibfnamefont {A.}~\bibnamefont
  {Deandrea}}, \bibinfo {author} {\bibfnamefont {N.}~\bibnamefont
  {Di~Bartolomeo}}, \bibinfo {author} {\bibfnamefont {R.}~\bibnamefont
  {Gatto}}, \bibinfo {author} {\bibfnamefont {F.}~\bibnamefont {Feruglio}}, \
  and\ \bibinfo {author} {\bibfnamefont {G.}~\bibnamefont {Nardulli}},\ }\href
  {\doibase 10.1016/S0370-1573(96)00027-0} {\bibfield  {journal} {\bibinfo
  {journal} {Phys. Rept.}\ }\textbf {\bibinfo {volume} {281}},\ \bibinfo
  {pages} {145} (\bibinfo {year} {1997})},\ \Eprint
  {http://arxiv.org/abs/hep-ph/9605342} {arXiv:hep-ph/9605342 [hep-ph]}
  \BibitemShut {NoStop}%
\bibitem [{\citenamefont {Weinberg}(1965)}]{Weinberg:1965zz}%
  \BibitemOpen
  \bibfield  {author} {\bibinfo {author} {\bibfnamefont {S.}~\bibnamefont
  {Weinberg}},\ }\href {\doibase 10.1103/PhysRev.137.B672} {\bibfield
  {journal} {\bibinfo  {journal} {Phys. Rev.}\ }\textbf {\bibinfo {volume}
  {137}},\ \bibinfo {pages} {B672} (\bibinfo {year} {1965})}\BibitemShut
  {NoStop}%
\bibitem [{\citenamefont {Braaten}\ and\ \citenamefont
  {Kusunoki}(2005)}]{Braaten:2005jj}%
  \BibitemOpen
  \bibfield  {author} {\bibinfo {author} {\bibfnamefont {E.}~\bibnamefont
  {Braaten}}\ and\ \bibinfo {author} {\bibfnamefont {M.}~\bibnamefont
  {Kusunoki}},\ }\href {\doibase 10.1103/PhysRevD.72.014012} {\bibfield
  {journal} {\bibinfo  {journal} {Phys. Rev.}\ }\textbf {\bibinfo {volume}
  {D72}},\ \bibinfo {pages} {014012} (\bibinfo {year} {2005})},\ \Eprint
  {http://arxiv.org/abs/hep-ph/0506087} {arXiv:hep-ph/0506087 [hep-ph]}
  \BibitemShut {NoStop}%
\bibitem [{\citenamefont {Guo}\ \emph {et~al.}(2015)\citenamefont {Guo},
  \citenamefont {Meißner}, \citenamefont {Wang},\ and\ \citenamefont
  {Yang}}]{Guo:2015umn}%
  \BibitemOpen
  \bibfield  {author} {\bibinfo {author} {\bibfnamefont {F.-K.}\ \bibnamefont
  {Guo}}, \bibinfo {author} {\bibfnamefont {U.-G.}\ \bibnamefont {Meißner}},
  \bibinfo {author} {\bibfnamefont {W.}~\bibnamefont {Wang}}, \ and\ \bibinfo
  {author} {\bibfnamefont {Z.}~\bibnamefont {Yang}},\ }\href {\doibase
  10.1103/PhysRevD.92.071502} {\bibfield  {journal} {\bibinfo  {journal} {Phys.
  Rev.}\ }\textbf {\bibinfo {volume} {D92}},\ \bibinfo {pages} {071502}
  (\bibinfo {year} {2015})},\ \Eprint {http://arxiv.org/abs/1507.04950}
  {arXiv:1507.04950 [hep-ph]} \BibitemShut {NoStop}%
\bibitem [{\citenamefont {Liu}\ \emph {et~al.}(2016)\citenamefont {Liu},
  \citenamefont {Wang},\ and\ \citenamefont {Zhao}}]{Liu:2015fea}%
  \BibitemOpen
  \bibfield  {author} {\bibinfo {author} {\bibfnamefont {X.-H.}\ \bibnamefont
  {Liu}}, \bibinfo {author} {\bibfnamefont {Q.}~\bibnamefont {Wang}}, \ and\
  \bibinfo {author} {\bibfnamefont {Q.}~\bibnamefont {Zhao}},\ }\href {\doibase
  10.1016/j.physletb.2016.03.089} {\bibfield  {journal} {\bibinfo  {journal}
  {Phys. Lett.}\ }\textbf {\bibinfo {volume} {B757}},\ \bibinfo {pages} {231}
  (\bibinfo {year} {2016})},\ \Eprint {http://arxiv.org/abs/1507.05359}
  {arXiv:1507.05359 [hep-ph]} \BibitemShut {NoStop}%
\bibitem [{\citenamefont {Guo}\ \emph {et~al.}(2016)\citenamefont {Guo},
  \citenamefont {Meißner}, \citenamefont {Nieves},\ and\ \citenamefont
  {Yang}}]{Guo:2016bkl}%
  \BibitemOpen
  \bibfield  {author} {\bibinfo {author} {\bibfnamefont {F.-K.}\ \bibnamefont
  {Guo}}, \bibinfo {author} {\bibfnamefont {U.-G.}\ \bibnamefont {Meißner}},
  \bibinfo {author} {\bibfnamefont {J.}~\bibnamefont {Nieves}}, \ and\ \bibinfo
  {author} {\bibfnamefont {Z.}~\bibnamefont {Yang}},\ }\href {\doibase
  10.1140/epja/i2016-16318-4} {\bibfield  {journal} {\bibinfo  {journal} {Eur.
  Phys. J.}\ }\textbf {\bibinfo {volume} {A52}},\ \bibinfo {pages} {318}
  (\bibinfo {year} {2016})},\ \Eprint {http://arxiv.org/abs/1605.05113}
  {arXiv:1605.05113 [hep-ph]} \BibitemShut {NoStop}%
\bibitem [{\citenamefont {Bayar}\ \emph {et~al.}(2016)\citenamefont {Bayar},
  \citenamefont {Aceti}, \citenamefont {Guo},\ and\ \citenamefont
  {Oset}}]{Bayar:2016ftu}%
  \BibitemOpen
  \bibfield  {author} {\bibinfo {author} {\bibfnamefont {M.}~\bibnamefont
  {Bayar}}, \bibinfo {author} {\bibfnamefont {F.}~\bibnamefont {Aceti}},
  \bibinfo {author} {\bibfnamefont {F.-K.}\ \bibnamefont {Guo}}, \ and\
  \bibinfo {author} {\bibfnamefont {E.}~\bibnamefont {Oset}},\ }\href {\doibase
  10.1103/PhysRevD.94.074039} {\bibfield  {journal} {\bibinfo  {journal} {Phys.
  Rev.}\ }\textbf {\bibinfo {volume} {D94}},\ \bibinfo {pages} {074039}
  (\bibinfo {year} {2016})},\ \Eprint {http://arxiv.org/abs/1609.04133}
  {arXiv:1609.04133 [hep-ph]} \BibitemShut {NoStop}%
\bibitem [{\citenamefont {Lu}\ \emph {et~al.}(2019)\citenamefont {Lu},
  \citenamefont {Geng},\ and\ \citenamefont {Valderrama}}]{Lu:2017dvm}%
  \BibitemOpen
  \bibfield  {author} {\bibinfo {author} {\bibfnamefont {J.-X.}\ \bibnamefont
  {Lu}}, \bibinfo {author} {\bibfnamefont {L.-S.}\ \bibnamefont {Geng}}, \ and\
  \bibinfo {author} {\bibfnamefont {M.~P.}\ \bibnamefont {Valderrama}},\ }\href
  {\doibase 10.1103/PhysRevD.99.074026} {\bibfield  {journal} {\bibinfo
  {journal} {Phys. Rev.}\ }\textbf {\bibinfo {volume} {D99}},\ \bibinfo {pages}
  {074026} (\bibinfo {year} {2019})},\ \Eprint
  {http://arxiv.org/abs/1706.02588} {arXiv:1706.02588 [hep-ph]} \BibitemShut
  {NoStop}%
\bibitem [{\citenamefont {Hu}\ and\ \citenamefont {Mehen}(2006)}]{Hu:2005gf}%
  \BibitemOpen
  \bibfield  {author} {\bibinfo {author} {\bibfnamefont {J.}~\bibnamefont
  {Hu}}\ and\ \bibinfo {author} {\bibfnamefont {T.}~\bibnamefont {Mehen}},\
  }\href {\doibase 10.1103/PhysRevD.73.054003} {\bibfield  {journal} {\bibinfo
  {journal} {Phys. Rev.}\ }\textbf {\bibinfo {volume} {D73}},\ \bibinfo {pages}
  {054003} (\bibinfo {year} {2006})},\ \Eprint
  {http://arxiv.org/abs/hep-ph/0511321} {arXiv:hep-ph/0511321 [hep-ph]}
  \BibitemShut {NoStop}%
\bibitem [{\citenamefont {Falk}\ and\ \citenamefont
  {Luke}(1992)}]{Falk:1992cx}%
  \BibitemOpen
  \bibfield  {author} {\bibinfo {author} {\bibfnamefont {A.~F.}\ \bibnamefont
  {Falk}}\ and\ \bibinfo {author} {\bibfnamefont {M.~E.}\ \bibnamefont
  {Luke}},\ }\href {\doibase 10.1016/0370-2693(92)90618-E} {\bibfield
  {journal} {\bibinfo  {journal} {Phys. Lett.}\ }\textbf {\bibinfo {volume}
  {B292}},\ \bibinfo {pages} {119} (\bibinfo {year} {1992})},\ \Eprint
  {http://arxiv.org/abs/hep-ph/9206241} {arXiv:hep-ph/9206241 [hep-ph]}
  \BibitemShut {NoStop}%
\bibitem [{\citenamefont {Cho}(1993)}]{Cho:1992cf}%
  \BibitemOpen
  \bibfield  {author} {\bibinfo {author} {\bibfnamefont {P.~L.}\ \bibnamefont
  {Cho}},\ }\href {\doibase 10.1016/0550-3213(94)90522-3,
  10.1016/0550-3213(93)90263-O} {\bibfield  {journal} {\bibinfo  {journal}
  {Nucl. Phys.}\ }\textbf {\bibinfo {volume} {B396}},\ \bibinfo {pages} {183}
  (\bibinfo {year} {1993})},\ \bibinfo {note} {[Erratum: Nucl.
  Phys.B421,683(1994)]},\ \Eprint {http://arxiv.org/abs/hep-ph/9208244}
  {arXiv:hep-ph/9208244 [hep-ph]} \BibitemShut {NoStop}%
\bibitem [{\citenamefont {Valderrama}(2012)}]{Valderrama:2012jv}%
  \BibitemOpen
  \bibfield  {author} {\bibinfo {author} {\bibfnamefont {M.~P.}\ \bibnamefont
  {Valderrama}},\ }\href {\doibase 10.1103/PhysRevD.85.114037} {\bibfield
  {journal} {\bibinfo  {journal} {Phys. Rev.}\ }\textbf {\bibinfo {volume}
  {D85}},\ \bibinfo {pages} {114037} (\bibinfo {year} {2012})},\ \Eprint
  {http://arxiv.org/abs/1204.2400} {arXiv:1204.2400 [hep-ph]} \BibitemShut
  {NoStop}%
\bibitem [{\citenamefont {Guo}\ \emph {et~al.}(2011)\citenamefont {Guo},
  \citenamefont {Hanhart}, \citenamefont {Li}, \citenamefont {Mei{\ss}ner},\
  and\ \citenamefont {Zhao}}]{Guo:2010ak}%
  \BibitemOpen
  \bibfield  {author} {\bibinfo {author} {\bibfnamefont {F.-K.}\ \bibnamefont
  {Guo}}, \bibinfo {author} {\bibfnamefont {C.}~\bibnamefont {Hanhart}},
  \bibinfo {author} {\bibfnamefont {G.}~\bibnamefont {Li}}, \bibinfo {author}
  {\bibfnamefont {U.-G.}\ \bibnamefont {Mei{\ss}ner}}, \ and\ \bibinfo {author}
  {\bibfnamefont {Q.}~\bibnamefont {Zhao}},\ }\href {\doibase
  10.1103/PhysRevD.83.034013} {\bibfield  {journal} {\bibinfo  {journal} {Phys.
  Rev.}\ }\textbf {\bibinfo {volume} {D83}},\ \bibinfo {pages} {034013}
  (\bibinfo {year} {2011})},\ \Eprint {http://arxiv.org/abs/1008.3632}
  {arXiv:1008.3632 [hep-ph]} \BibitemShut {NoStop}%
\end{thebibliography}%

\end{document}